\documentclass[preprint,12pt,authoryear]{elsarticle}

\usepackage{macro}

\begin{document}

\begin{frontmatter}

\title{Modelling the passive and active response of skeletal muscles within the adapted Voigt representation framework}

\author[1,2]{Sara Galasso}
            \ead{sara.galasso@unipd.it}
\author[1,2]{Giulio G. Giusteri\texorpdfstring{\corref{cor1}}{}}
            \ead{giulio.giusteri@unipd.it}
            \cortext[cor1]{Corresponding author}

\affiliation[1]{
            organization={Dipartimento di Matematica ``Tullio Levi-Civita'', Università degli Studi di Padova},
            addressline={Via Trieste 63}, 
            city={Padova},
            postcode={35121}, 
            country={Italy}}
\affiliation[2]{
            organization={Gruppo Nazionale per la Fisica Matematica, Istituto Nazionale di Alta Matematica ``Francesco Severi''},
            city={Sezione di Padova},
            country={Italy}}

\begin{abstract}
We present a constitutive model for the passive and active response of skeletal muscles.
At variance with more classical approaches, the model is developed exploiting adapted Voigt representations of strain and stress tensors within the context of nonlinear Cauchy elasticity.
This framework allows us to identify non-trivial stress--strain relations in a rather direct way from experimental data, enhancing the mechanical interpretability of the material functions that describe the tissue response and obtaining additional insight on the distinct role of the contractile fibres and of the surrounding extracellular matrix. 
We propose a two-material model, with an additive splitting of the stress contributions, in which only one component depends on an activation parameter. The constitutive model for the passive behaviour satisfactorily predicts the nonlinear stress response to elongation at different relative orientations with respect to the fibre direction and highlights the dominant role of the extracellular matrix. The active term, essentially determined by the mechanics of the contractile fibres, captures well the isometric stress response of tetanised tissue through the prescription of an along-fibre active strain that also depends on the matrix deformation.
\end{abstract}

\begin{keyword}
biomechanics \sep skeletal muscles \sep anisotropic nonlinear elasticity \sep active strain \sep adapted tensorial basis
\MSC[2020] 74L15 \sep 74B20 \sep 74E10

\end{keyword}

\end{frontmatter}

\section{Introduction}
The recent development of active matter has stimulated a re-examination of constitutive modelling strategies in continuum mechanics. Active solids, such as biological tissues endowed with internal or externally supplied energy sources, do not naturally fit within classical variational hyperelastic frameworks. In this context, the setting of Cauchy elasticity, recently revisited by \cite{yavari2025nonlinear}, appears particularly flexible, as it allows one to constitutively prescribe the stress response directly, without requiring the existence of an underlying energy function.
This perspective aligns naturally with the emerging data-driven paradigm in mechanics, in which constitutive relations are inferred directly from experimental observations rather than obtained by calibrating parameters within a pre-assigned model \citep{montans2019data,tikenogullari2023data}. 
With the goal of facilitating this modelling procedure, \cite{galasso2025adapted} introduced a theoretical framework for anisotropic nonlinear elasticity based on the decomposition of strain and stress tensors on a tensorial basis adapted to the local anisotropy of the material. Their approach consists in constitutively prescribing the independent components of the stress (rather than an energy function), inferring them directly from experimental data. Constitutive hypotheses, such as frame indifference, material symmetries or incompressibility, are granted by suitably choosing the adapted tensorial basis used for the decomposition.
A notable feature of this approach is that the resulting material functions---which characterise the response of the material---retain a clear mechanical interpretation, being directly associated with experimentally measurable stress components. In the present work, we exploit this framework to model the passive and active quasi-static mechanical response of skeletal muscle tissue from uniaxial tensile data, thereby constructing a constitutive model that also provides significant biophysical insight.

The mechanical behaviour of skeletal muscle is the macroscopic manifestation of a highly organised and hierarchical biological structure, in which each constituent contributes to a specific physiological function \citep{lieber2011skeletal,mukund2020skeletal}. The primary roles of skeletal muscles include force production and movement, achieved through active contraction triggered by electrochemical stimulation. At the microscale, force is generated within sarcomeres, the fundamental contractile units composed of cross-bridging actin and myosin filaments. Sarcomeres are arranged in series to form myofibrils, which in turn assemble into elongated cells known as muscle fibres. Muscle fibres are embedded in and mechanically coupled through an extracellular matrix, a collagen-rich network organised across multiple structural levels (endomysium, perimysium, and epimysium), forming a complex three-dimensional architecture. As a consequence of this multilevel organisation, skeletal muscle exhibits an intrinsically composite mechanical behaviour: active force generation arises from the actomyosin interactions within the fibres, whereas passive elasticity and load redistribution are primarily governed by the extracellular matrix and structural proteins, such as titin. A detailed account of skeletal muscle anatomy and physiology is given, for instance, by \cite{oatis2009kinesiology} and \cite{luis2021mechanobiology}.

The complex multiscale organisation of the muscular system poses significant challenges for the experimental characterisation and for the interpretation of mechanical responses. Measured stress–strain relations are highly sensitive to the experimental protocol, including activation conditions (twitch versus tetanic stimulation), loading mode (isometric, isotonic, eccentric, or concentric contractions), strain rate, preconditioning history, and specimen preparation (single fibre, fibre bundle, fascicle, or whole muscle). Consequently, quantitative comparisons across studies are often non-trivial, and the extraction of intrinsic material properties from experimental data remains a delicate task \citep{binder2021review,kohn2021direction,lieber2023extracellular}. Moreover, most experiments provide limited access to full three-dimensional stress states, complicating the identification of constitutive models within a finite-strain framework \citep{giantesio2019comparison}. The basic mechanical properties are typically investigated through isometric force–length tests and quasi-static tensile experiments, conducted at different structural levels (isolated sarcomeres, single fibres, fascicles, or whole muscles). While active force generation scales approximately linearly from sarcomeres to higher hierarchical levels---consistent with the serial and parallel arrangement of contractile units---, passive stiffness does not exhibit a comparable scaling behaviour \citep{winter2011whole,ward2020scaling}. At larger scales, the passive response becomes increasingly dominated by the extracellular matrix, leading to marked nonlinearities and anisotropy at the tissue level \citep{ogden2003nonlinear,kohn2021direction}.
More advanced experimental protocols, including cyclic tests and eccentric–concentric contractions, reveal marked history dependence and dissipative effects \citep{ogden2003nonlinear}.

At the cellular level, the fundamental biophysical mechanisms underlying active force generation are relatively well established, beginning with the sliding filament and cross-bridge theories, which provide a mechanistic description of actomyosin interactions. In contrast, the mechanical behaviour of skeletal muscle at the tissue scale remains only partially understood. The collective response emerging from the interaction between fibres, extracellular matrix, and neural activation involves strong nonlinearities, anisotropy, history dependence, and multiscale coupling. As a consequence, existing mathematical models often struggle to reconcile physiological fidelity with mechanical consistency and computational tractability, and their predictive capabilities outside calibrated experimental settings remain limited. A rigorous mechanical description of skeletal muscle at the tissue scale naturally calls for a continuum–mechanical framework capable of linking microscopic mechanisms to macroscopic observables \citep{epstein1998theoretical,epstein2012elements,dao2018systematic}.
Several modelling strategies have been proposed to bridge the gap between microscopic structure and macroscopic response \citep{chapelle2012energy,miller2021microstructurally,loumeaud2024multiscale}. Multiscale approaches attempt to explicitly link sarcomere-level dynamics to tissue-level behaviour through scale-transition procedures. Homogenisation techniques have been employed to derive macroscopic constitutive laws from idealised microstructural representations of fibres embedded in an extracellular matrix, providing a systematic framework to upscale structural information while preserving mechanical consistency \citep{spyrou2016homogenization}. Complementary to these efforts, multiphysics models have been developed to couple mechanical deformation with electrochemical processes, prescribing physiologically based activation kinetics \citep{roehrle2012physiologically,roehrle2019multiscale,stefanati2020mathematical}. Although these frameworks provide valuable insight, their complexity often limits their applicability in large-scale simulations or in contexts where only macroscopic observables are available.

Within the purely mechanical setting, the dominant paradigm for three-dimensional constitutive modelling of the static behaviour of the tissue is finite hyperelasticity. Passive response is described through strain-energy functions depending on generalised strain invariants associated with preferred fibre directions, subject to requirements such as polyconvexity and coercivity to ensure well-posedness \citep{schroeder2003invariant,ogden2003nonlinear,wollner2025search}. \cite{chagnon2014hyperelastic} and \cite{dao2018systematic} review many choices of hyperelastic energy functions, while other studies account for the complex structure of the extracellular matrix \citep{salvatore2010nonlinear,bleiler2019micro}.
Activation is then incorporated through several distinct strategies. One class extends the dependency of the strain energy to additional deformation-dependent internal variables, which control microstructural changes due to the active contraction of the fibres \citep{itskov2009universal,ehret2011continuum,giantesio2017strain,stefanati2020mathematical}. More in general, the active strain approach assumes that activation modifies the stress-free state of the material, and this is attained by a multiplicative decomposition of the deformation gradient, where the active component is prescribed constitutively \citep{taber2000modeling,nardinocchi2007active,ambrosi2011perspectives,devita2017mathematical}. 
The strain energy depends then solely on the elastic component, thereby preserving the passive functional form \citep{giantesio2018loss}. When viewed as a two-phase or fibre-reinforced composite, the muscle is typically modelled by a strain-energy function which additively decouples into a purely passive (frequently isotropic) component and an active (transversely isotropic) one---an approach termed mixture active strain by \cite{riccobelli2019activation}. 
Alternatively, the active stress approach models activation as an externally generated stress contribution which linearly superposes to the passive stress \citep{blemker2005model,roehrle2008bridging,wheatley2018modeling}. A comprehensive overview of active modelling strategies is provided in \cite{goriely2017five}, while \cite{engelhardt2025constitutive} offer a recent comparative review. Further comparisons and features of these methods are discussed, for instance, by \cite{ambrosi2011active,giantesio2019comparison,klotz2021physiology,giantesio2024modeling}.

Adopting a mixed approach, we propose a two-material model, with an additive splitting of the stress contributions, in which only one component depends on a non-trivial active strain. The \emph{matrix} component is responsible for the passive behaviour and its response to strain is independent of the activation state. The \emph{fibre} component contributes to the mechanical response when an activation control parameter is switched on and requires the prescription of an elastically-relaxed state dependent on the matrix strain. The additional stress produced by the tissue, when active, is then viewed as a reorganisation of the microscopic structure, which, however, interests one component of the material only.
While additive decompositions of passive and active stresses are widely employed in muscle models, the present work differs in its objective: rather than introducing a new activation mechanism, it develops a constitutive representation within the adapted Voigt framework that provides a transparent parameterisation of passive and active stress contributions.

The article has the following structure. Section \ref{sec:setting} defines the theoretical framework, following \cite{galasso2025adapted}. In Section \ref{sec:constitutive_modelling}, we specify the experimental data upon which we construct our model, and we introduce the two-material constitutive hypothesis. The two components of the stress are analysed in the following sections. Specifically, Section \ref{sec:passive_model} deals with the mechanical response of the passive tissue, while Section \ref{sec:active_model} is devoted to the active response. Conclusions are presented in Section \ref{sec:conclusions}.

\section{Theoretical framework}
\label{sec:setting}
We start by constructing a theoretical framework, in the setting of Cauchy nonlinear elasticity, which exploits the biophysical structure of the system we aim to study. The motivation and the general setting for this formalism have been discussed by \cite{galasso2025adapted}. Specifically, in such a scheme, the structure of the constitutive law is designed for a data-driven approach, which enables the prescription of material functions directly from stress--strain measurements. Each degree of freedom has a clear and objective mechanical meaning, and the independence of stress components facilitates the physical interpretation of distinct elastic response contributions. To obtain a convenient representation of the constitutive relation which features incompressibility and transverse isotropy, we introduce the following local adapted tensorial basis, which allows us to work with adapted Voigt representations of strain and stress tensors.

\subsection{Adapted bases}
We fix, at any material point $X$ in the reference configuration $\mathcal{B}\subset\RR^3$, an adapted basis $\mathcal{L}=\{\*l_1,\*l_2,\*l_3\}$ for the tangent space $T_X\mathcal{B}$, such that $\mathcal{L}$ is orthonormal with respect to the Euclidean scalar product and aligned with the local orientation of the fibres in $X$, namely $\*l_i\cdot\*l_j=\delta_{ij}$ and $\*l_1\otimes\*l_1$ identifies the local along-fibre direction. We define the material tensorial basis $\mathcal{Z}=\{\+Z_1,\dots,\+Z_6\}$ for the vector space $\text{Sym}(T_X\mathcal{B})$ of second-order symmetric tensors at $X\in\mathcal{B}$ as follows:
\begin{align*}
\+Z_1&=\sqrt{\frac23}\left(\*l_1\otimes\*l_1-\frac{\*l_2\otimes\*l_2}{2}-\frac{\*l_3\otimes\*l_3}{2}\right), 
&\+Z_4&=\frac{\*l_1\otimes\*l_2+\*l_2\otimes\*l_1}{\sqrt{2}},\\
\+Z_2&=\frac{\*l_2\otimes\*l_2-\*l_3\otimes\*l_3}{\sqrt{2}},
&\+Z_5&=\frac{\*l_2\otimes\*l_3+\*l_3\otimes\*l_2}{\sqrt{2}},\\
\+Z_3&=\frac{\*l_1\otimes\*l_3+\*l_3\otimes\*l_1}{\sqrt{2}},
&\+Z_6&=\frac{\*l_1\otimes\*l_1+\*l_2\otimes\*l_2+\*l_3\otimes\*l_3}{\sqrt{3}}.
\end{align*}
This basis satisfies the orthonormality condition $\+Z_i:\+Z_j=\delta_{ij}$, with respect to the tensor scalar product $\+A:\+B=\tr(\+A^{\+T}\+B)$. Furthermore, $\tr{\+Z_i}=0$ for $i\neq6$ and $\tr{\+Z_6}=\sqrt{3}$. The introduction of such a tensorial basis allows us to work with (vectorial) Voigt representations of the strain and the stress, which are, in fact, adapted to the transverse isotropy and incompressibility of the system we are going to analyse.

\subsection{Multiplicative decompositions}
Let $\+F:T\mathcal{B}\to T\varphi(\mathcal{B})$ be the deformation gradient of a configuration $\varphi:\mathcal{B}\to\RR^3$, namely $\+F(X)=\nabla\varphi(X)$ for any material point $X\in\mathcal{B}$. We exploit the Kröner--Lee multiplicative decomposition of the deformation gradient (see~\cite{yavari2023decomposition} for a detailed discussion), by which we assume that locally there is a decoupling into an elastic component $\+F_{\el}$ and an elastically-relaxed one $\+F_0$, and it reads $\+F=\+F_{\el}\+F_{0}$. The field $\+F_0$ is a material tensor field, while $\+F_{\el}$ is a two-point tensor field that maps material vectors into spatial ones. Unlike the deformation gradient $\+F$, the tensor fields $\+F_{\el}$ and $\+F_0$ might be geometrically incompatible, in the sense that they are not necessarily gradient fields. This means that the compatibility of the relaxed configuration with the Euclidean metric of the ambient space is restored by the elastic component $\+F_\el=\+F\+F_0^{-1}$, which induces elastic stresses in the body. In particular, a (possibly non-constant) term $\+F_0$ different from the identity implies that the material is anelastic in the sense that its configurations could be elastically stressed even in the absence of external forces \citep{eckart1948anelasticity}. In the context of active matter, the term $\+F_0$ is associated with a remodelling of the microscopic structure of the material due to activation, which converts electrochemical energy into mechanical work \citep{nardinocchi2007active}. For muscles, this process permits the tissue to contract in the absence of external loads.
We assume that both the elastically-relaxed and the elastically-stressed contributions do not produce changes in the volume of the body, hence we have the incompressibility conditions $\det\+F_0=1$ and $\det\+F_\el=1$, ensuring that $J=\det\+F=1$. In the following, we exploit also the polar decomposition of the elastic component of the deformation gradient: $\+F_\el=\+R_\el\+U_\el$, where $\+R_\el$, orthogonal, is a rotation and $\+U_\el$, symmetric and positive-definite, is the right stretch tensor.

\subsection{Adapted Voigt representations}
As a measure of elastic strain, with respect to the reference configuration $\mathcal{B}$, we consider the right Cauchy--Green strain tensor of the elastic component $\+C_{\el}=\+F_{\el}^{\+T}\+F_{\el}$, related to the right stretch tensor by $\+C_\el=\+U_\el^2$. By definition, $\+C_\el(X)$ belongs to the space $\text{Sym}^+(T_X\mathcal{B})$ of second-order symmetric positive-definite tensors at $X$. Its tensorial logarithm, $\log:\text{Sym}^+(T\mathcal{B})\to\text{Sym}(T\mathcal{B})$, referred to as Hencky strain, provides us a map from the multiplicative group of deformation gradients to its Lie algebra, making it the natural choice of strain to enter the constitutive relation \citep{latorre2014interpretation,neff2016geometry,giusteri2024evolution}. We hence decompose the elastic strain $\log\+C_\el$ on the material basis $\mathcal{Z}$:
\[
  \frac12\log\+C_{\el}=\log\+U_{\el}=\sum_{k=1}^6\lambda_k\,\+Z_k,
\]
with $\l_k=\log\+U_{\el}:\+Z_k\in\RR$, for $k=1,\dots,6$. In particular, at the identity $\+U_{\el}=\id_3$, all the components $\l_k$ vanish.
From the incompressibility constraint, it follows that the volumetric component of the strain vanishes, namely $\lambda_6=0$. Indeed, the condition $\det\+F_{\el}=1$ gives $\tr(\log\+C_{\el})=0$, since $\det\exp(\log\+C_{\el})=\exp\tr(\log\+C_{\el})$. In this case, the coordinates of the elastic strain $\log\+U_\el$ on the tensorial basis are given by
\begin{align*}
\l_1&=\sqrt{\frac32}\,\*l_1\cdot \log\+U_{\el}\,\*l_1,
& \l_4&=\sqrt{2}\,\*l_2\cdot\log\+U_{\el}\,\*l_1,\\
\l_2&=\frac{\*l_2\cdot\log\+U_{\el}\,\*l_2-\*l_3\cdot\log\+U_{\el}\,\*l_3}{\sqrt{2}},
& \l_5&=\sqrt{2}\,\*l_2\cdot\log\+U_{\el}\,\*l_3,\\
\l_3&=\sqrt{2}\,\*l_3\cdot\log\+U_{\el}\,\*l_1,
& \l_6&=0.
\end{align*}
The orthonormality of the tensorial basis implies that each component $\l_k$ represents an independent contribution to the strain and their mechanical meaning, independently of the reference frame, is the following: $\lambda_1$ is the distortional part of along-fibre stretch ($>0$ elongation, $<0$ compression), $\lambda_2$ gives the difference between cross-fibre stretches, $\lambda_3$ and $\lambda_4$ represent along-fibre pure shears, and $\lambda_5$ is the cross-fibre pure shear.
The coordinate vector $\*\l=(\l_1,\dots,\l_5)\in\RR^5$ is thus an adapted and objective Voigt representation of the distortional part of the elastic strain $\log\+U_{\el}$.
\\

As a measure of stress on the body in the reference configuration we adopt the second Piola--Kirchhoff stress tensor $\+S_{\el}$. Since $\+S_{\el}(X)\in\text{Sym}(T_X\mathcal{B})$, we decompose it on the basis $\mathcal{Z}$:
\[
  \+S_{\el}=\sum_{k=1}^6\sigma_k\,\+Z_k
\]
with components $\sigma_k=\+S_\el:\+Z_k\in\RR$, $k=1,\dots,6$, representing independent stress response terms. In particular, 
$\sqrt{6}\sigma_1=2\*l_1\cdot\+S_{\el}\,\*l_1-\*l_2\cdot\+S_{\el}\,\*l_2-\*l_3\cdot\+S_{\el}\,\*l_3$ gives the deviatoric along-fibre normal stress, $\sqrt{2}\sigma_2=\*l_2\cdot\+S_{\el}\,\*l_2-\*l_3\cdot\+S_{\el}\,\*l_3$ the difference between the cross-fibre normal stresses, $\sigma_3=\sqrt{2}\*l_3\cdot\+S_{\el}\,\*l_1$ and $\sigma_4=\sqrt{2}\*l_2\cdot\+S_{\el}\,\*l_1$ the along-fibre pure shear stresses, $\sigma_5=\sqrt{2}\*l_2\cdot\+S_{\el}\,\*l_3$ the pure shear stress in the cross-fibre plane, and $\sqrt{3}\sigma_6=\*l_1\cdot\+S_{\el}\,\*l_1+\*l_2\cdot\+S_{\el}\,\*l_2+\*l_3\cdot\+S_{\el}\,\*l_3$ the spherical part of the stress.
In the incompressible case, the kinematic constraint $\det\+F=1$ is enforced by a Lagrange multipliers $p$---a scalar field representing the hydrostatic pressure---defined as the spherical part of the Cauchy stress $\+T_{\el}$: $p=-\tfrac13\tr\+T_{\el}$, where $\+T_{\el}=J^{-1}\+F\+S_{\el}\+F^{\+T}$. This relation translates in a linear constraint on the stress components $\sigma_i$'s, given by
\[
  \sum_{i=1}^6\+\Lambda_{6i}\sigma_i+\sqrt{3}p=0,
\]
where $\+\Lambda_{ij}=J^{-1}\+U_{\el}\+Z_j\+U_{\el}^{\+T}:\+Z_i$ and the hydrostatic pressure $p$ is to be determined from the momentum balance equation and the prescribed boundary conditions of the problem. Hence, only five out of the six stress components are mutually independent. In particular, we can express $\sigma_6=-\+\Lambda_{66}^{-1}(\sqrt{3}p+\sum_{i=1}^5\+\Lambda_{6i}\sigma_i)$.
We then define the coordinate vector $\*\sigma=(\sigma_1,\dots,\sigma_5)\in\RR^5$, which gives an adapted and objective Voigt representation of the deviatoric part of the elastic stress $\+S_{\el}$.

\subsection{Constitutive framework}
For a homogeneous Cauchy elastic material, the mechanical response is a function of the strain only. In particular, the elastic stress depends solely on the elastic strain, namely the constitutive equation has the form $\+S_{\el}=\hat{\+S}_{\el}(\log\+U_{\el})$, for some tensor-valued map $\hat{\+S}_\el:\text{Sym}(T_X\mathcal{B})\to\text{Sym}(T_X\mathcal{B})$ such that $\hat{\+S}_\el(\+0)=\+0$.
In our setting, this stress--strain relation, for an incompressible elastic material, is represented by a nonlinear vector field on $\RR^6$, $\hat{\*\sigma}:\RR^5\to\RR^5$, $\*\sigma=\hat{\*\sigma}(\*\l)$, satisfying $\hat{\*\sigma}(\*0)=\*0$. 
Alongside $\hat{\*\sigma}$, we need to postulate also the form of $\+F_0$, which might itself depend on the strain.
For a generic anisotropic material, the material functions are determined by constitutively prescribing the functional dependence of the five independent stress components on the five strains and the relaxed state associated with activation.

As discussed by~\cite{galasso2025adapted}, whenever the response of the material manifests some invariance properties under a class of deformations, namely the system has a material symmetry, the vector field $\hat{\*\sigma}$ transforms covariantly under the linear action of the corresponding symmetry group. As a consequence, the number of independent strain invariants (the $\l_k$'s, $k=1,\dots,6$) diminishes and $\hat{\*\sigma}$ admits an irreducible representation. We are interested in the case of transverse isotropy, that is in material responses which are invariant under rotations about the preferred direction along $\*l_1$.  Explicitly, $\hat{\*\sigma}(\+\Omega \*\l)=\+\Omega\hat{\*\sigma}(\*\l)$ with $\+\Omega_{ij}=\+Q^{\+T}\+Z_j\+Q:\+Z_i$, where $\+Q$ is any rotation about $\*l_1$ or a reflection in the corresponding orthogonal plane \citep{galasso2025adapted}.
For a transversely isotropic material, the strain has five independent invariants
and the constitutive relation admits an irreducible representation given by the sum of six independent terms with coefficients that are functions of the invariants \citep{delpiero1998representation}. 
The kinematic constraint of incompressibility reduces the number of invariants to four, since $\tr(\log\+U_\el)=0$, and the number of independent terms to five, since one coefficient of the decomposition is always expressible in terms of the remaining ones and the pressure.
We choose the invariants as follows: $\hat\rho_1(\*\l)=\lambda_1$, which is the along-fibre stretch, $\hat\rho_2(\*\l)=\lambda_2^2+\lambda_5^2$, which accounts for the magnitude of the cross-fibre deformations (both normal and tangential), $\hat\rho_3(\*\l)=\lambda_3^2+\lambda_4^2$, which quantifies the magnitude of the along-fibre tangential contributions to the strain, and $\hat\rho_4(\*\l)=2\lambda_3\lambda_4\lambda_5+\lambda_2(\lambda_3^2-\lambda_4^2)$. A direct computation allows to check that indeed $\hat{\rho}_i(\+\Omega\*\l)=\hat{\rho}_i(\*\l)$ for $i=1,\dots,4$. We shall also use the following functionally dependent invariant: $\hat{\rho}(\*\l)=\sqrt{\hat{\rho}_1^2(\*\l)+\hat{\rho}_2(\*\l)+\hat{\rho}_3(\*\l)}=(\sum_{k=1}^6\l_k^2)^{1/2}$, which measures the overall local strain. 
For the transversely isotropic and incompressible material, an irreducible representation for the constitutive relation reads then
\beq{eq:stress_incompressible_transverse}
      \hat{\*\sigma}(\*\l)=
      c_1\!\mx{\l_1\\0\\0\\0\\0}
      +c_2\!\mx{0\\\l_2\\0\\0\\\l_5}
      +c_3\!\mx{0\\0\\\l_3\\\l_4\\0}
      +c_4\!\mx{0\\0\\\l_4\l_5-\l_2\l_3\\\l_2\l_4+\l_3\l_5\\0}
      +c_5\!\mx{0\\\l_4^2-\l_3^2\\0\\0\\2\l_3\l_4}
\eeq
with $c_i=\hat{c}_i(\rho_1,\dots,\rho_4)$, $i=1,\dots,5$, with $\rho_k=\hat{\rho}_k(\*\l)$ for $k=1,\dots,4$. 
The constitutive response of the material is encoded in the five material functions $\hat{c}_i:\RR^4\to\RR$ which depend nonlinearly on four invariants.

\section{Experimental data sets and constitutive modelling}
\label{sec:constitutive_modelling}
Passive and active mechanical behaviours of skeletal muscles exhibit distinct characteristics, which reflect structural components of the tissue having different physiological roles \citep{mukund2020skeletal}. In particular, in the absence of activation by electrochemical stimuli, the majority of the load is borne by the extracellular matrix, with a smaller contribution from the fibres (specifically, from the protein titin). On the contrary, the fibres are the only actively contractible component (involving, specifically, myosin and actin proteins). We want our model to capture these distinct contributions and account for the microstructure of the material in our macroscale description. 

We restrict our investigation to quasi-static conditions, without accounting for rate-dependent effects. Furthermore, the active response is considered under isometric conditions in tetanic regime, focusing on the sustained force generated by muscle stimulation at a fixed length rather than on the transient dynamics of activation and contraction.

\subsection{Uniaxial quasi-static tensile tests}
\begin{figure}[htb]
\centering
\includegraphics[width=0.495\textwidth]{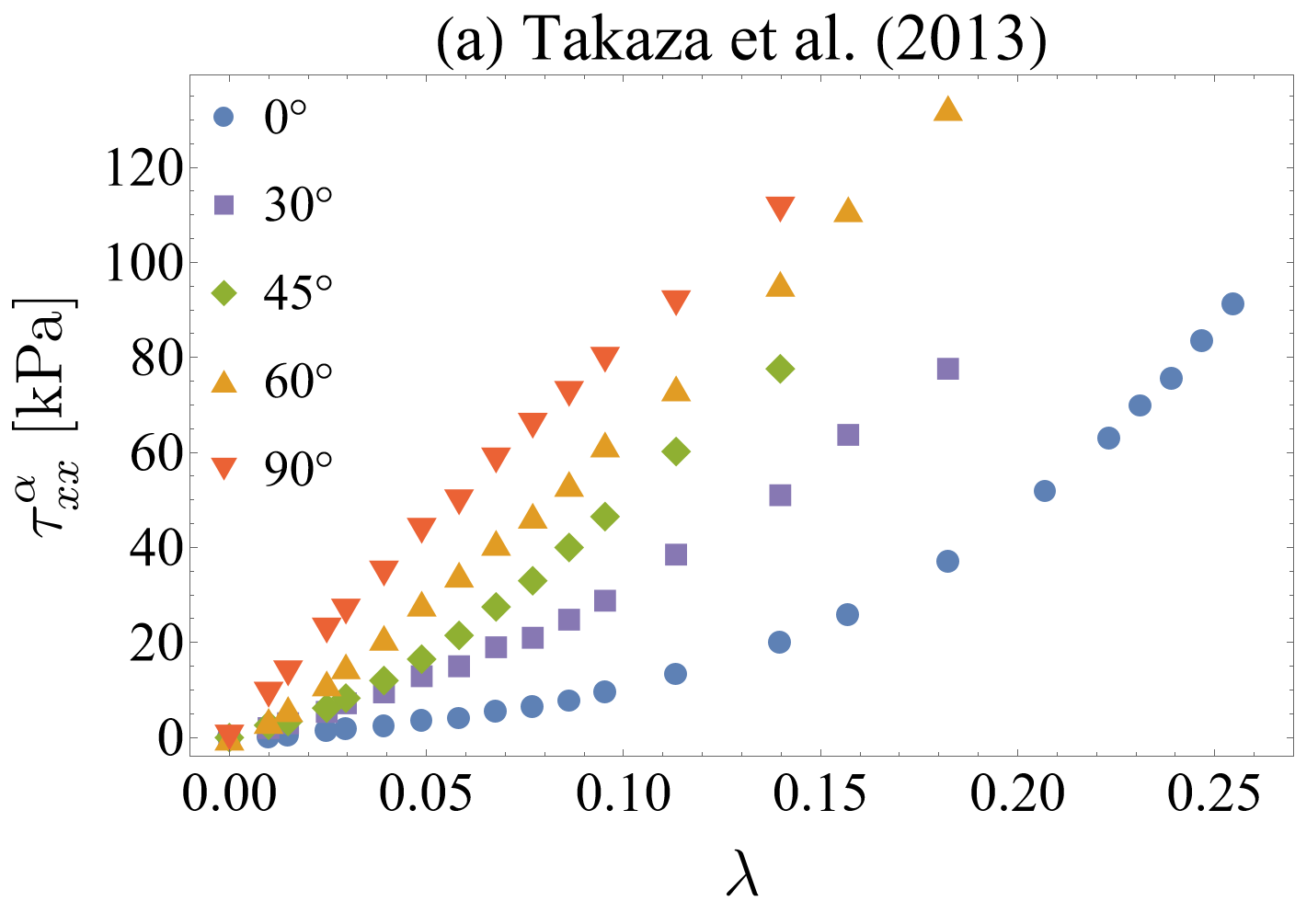}
\includegraphics[width=0.495\textwidth]{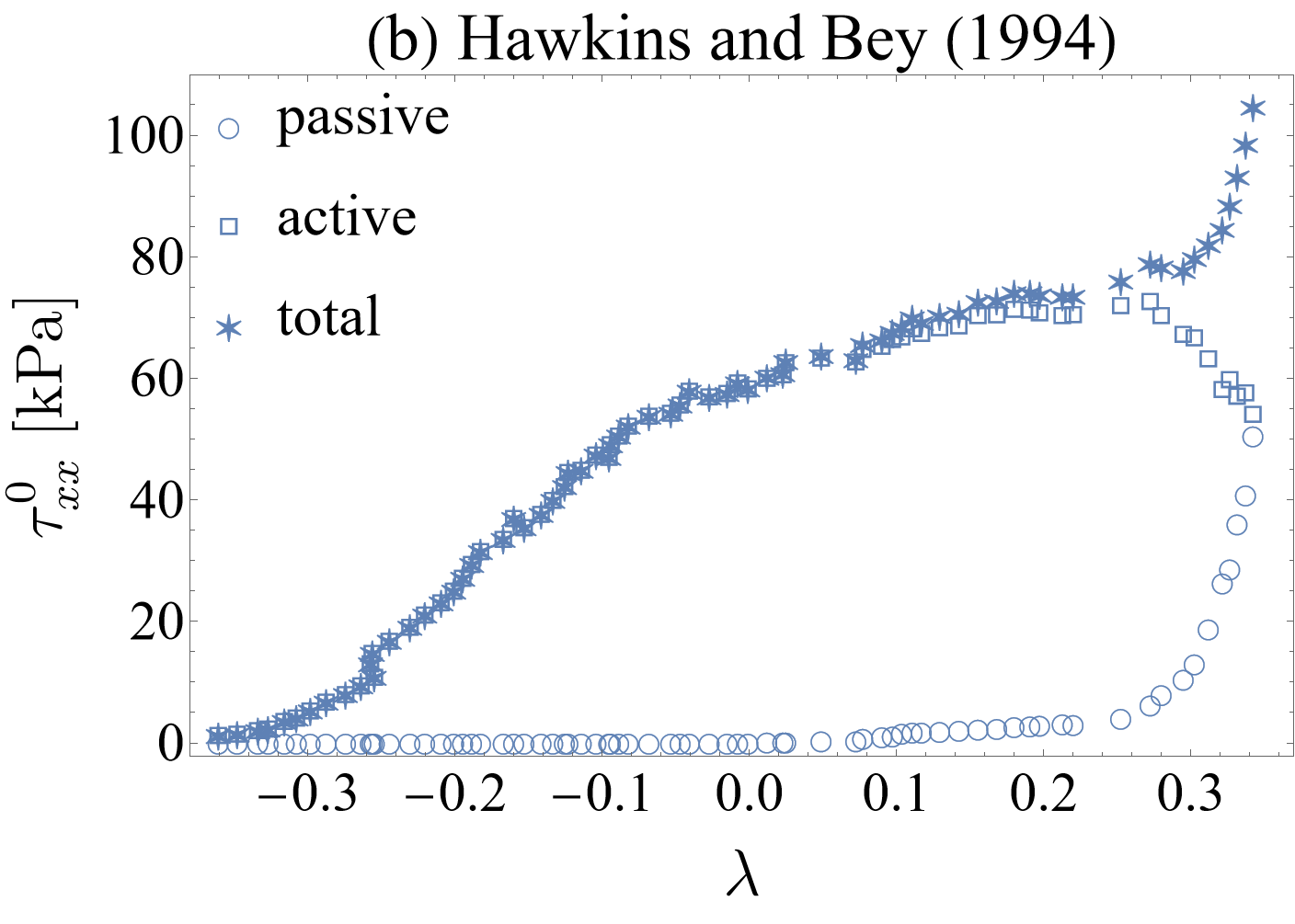}
\caption{Experimental data sets used for the construction of the model: uniaxial tensile tests performed at quasi-static rate on (a) passive~\cite{takaza2013anisotropic} and (b) tetanised~\cite{hawkins1994comprehensive} samples of skeletal muscle tissue. The reported data collect the component of the Cauchy stress tensor $\tau_{xx}^\a$ as a function of the Hencky stretch ratio $\l=\log\left(\tfrac{L}{L_0}\right)$, measured along a direction inclined at an angle $\a$ with respect to the fibre orientation.}
\label{fig:data}
\end{figure}

We construct our model for the elastic response of the skeletal muscle tissue on the experimental data by~\cite{takaza2013anisotropic}, for the passive response, and by~\cite{hawkins1994comprehensive}, for the active behaviour. In the latter case, the experimental protocol considers fused tetanic stimulation under isometric loading, which is, the tissue is electrically stimulated to maintain a sustained activated state, while keeping its length constant.
These experimental studies provide stress--strain pairs collected from uniaxial tensile tests that are performed on laterally unconstrained tissue samples---at a constant quasi-static strain rate. The procedures are optimised to investigate the elastic response of the material (e.g., with negligible viscous effects). Forces and elongations are measured and converted into stresses and strains by normalising each of them with the cross-sectional area and the reference length, respectively. Specifically, the local deformation, assumed homogeneous, is quantified by the Hencky stretch ratio $\l=\log\left(\tfrac{L}{L_0}\right)$, where $L_0$ is a set reference length and $L$ the loaded length. The normal component of the Cauchy stress in the direction of stretching is computed as the ratio between the applied load and the cross-sectional area in the bulk. Fixed the lab reference frame $\*e_x=(1,0,0),\*e_y=(0,1,0),\*e_z=(0,0,1)$, with $\*e_x$ the direction of stretching, the direction of the fibres, in the $xy$-plane, forms generically an angle $\alpha$ from the $x$ axis. The adapted vectorial basis $\mathcal{L}$ is then given by $\*l_1(\a)=\*e_x\cos\alpha+\*e_y\sin\alpha$, $\*l_2(\a)=-\*e_x\sin\alpha+\*e_y\cos\alpha$, and $\*l_3(\a)=\*e_z$. Accordingly, the elements of the adapted tensorial basis $\+Z_k(\a)$, $k=1,\dots,6$, can be explicitly computed. 
The deformation gradient of a homogeneous isochoric uniaxial deformation along the $x$-direction is represented, in the chosen coordinate system, by the matrix 
\[
  \+F=\mx{e^{\l}&0&0\\0&e^{-\l/2}&0\\0&0&e^{-\l/2}}.
\]
The $xx$-component of the Cauchy stress tensor $\+T_\el$, measured as a function of $\l$ at each $\a$, is obtained by projection along $\*e_x\otimes\*e_x$ as
$\tau_{xx}^\a=\*e_x\cdot\+T_\el\*e_x$.
The measurements by~\cite{takaza2013anisotropic} on passive tissue consist of tensile tests performed at various directions with respect to the mean fibre orientation: along, perpendicularly, and at the intermediate angles $30^\circ,45^\circ,60^\circ$ with respect to the fibre direction, see Figure~\ref{fig:data}(a).
\cite{hawkins1994comprehensive} tested both the passive and the active responses by stretching along the fibre direction, see Figure~\ref{fig:data}(b).

In a material picture, we explicit the components of the stress at different fibre orientation, $\sigma_k^\a$, by substituting $\+T_\el=\+F\left(\sum_{k}\sigma_k^\a\+Z_k(\a)\right)\+F^{\+T}$, with the constraint $p=-\tfrac13\tr\+T_{\el}$.
For static homogeneous deformations, the hydrostatic pressure is spatially constant, and its value is fixed by imposing the boundary conditions. As is common in uniaxial tests on unconstrained samples, we take $p\equiv0$.

Hence, the measured component of the Cauchy stress is related to the components of the second Piola--Kirchhoff stress by
\beq{eq:measured_stress}
  \tau_{xx}^\a=\sqrt{\frac{3}{8}}\frac{1}{2e^{-2\l}+e^\l}\left[(1+3\cos(2\a))\sigma_1^\a+\sqrt{3}(1-\cos(2\a))\sigma_2^\a-2\sqrt{3}\sin(2\a)\sigma_4^\a\right].
\eeq
The measurements proposed in the literature activate one or more stress components by changing the direction of the stretching with respect to the orientation of the fibres. In particular, the chosen data sets provide us with some information on the normal stresses $\sigma_1$, $\sigma_2$, and on the shear stress $\sigma_4$.

\subsection{Two-material model}
\label{sbsec:two-material_model}
We propose a two-material model, distinguishing the active component, which we refer to as `\emph{fibre}', from the passive one, which we refer to as `\emph{matrix}'. The distinction between the two components is intended to be understood in terms of different mechanical responses at the macroscale, rather than being directly interpreted as biophysically distinct structures at the cellular level. We assume that both materials are transversely isotropic and incompressible. Quantities referred to the matrix component are labelled with a sub- (or super-) script `$\ecm$', while those referred to the fibre with `$\fibre$'.

We introduce the following notation for the multiplicative decomposition of the deformation gradient $\+F$: we denote by $\+F_{0,\ecm}$ and $\+F_{0,\fibre}$ the material tensor fields for the matrix and fibre components, respectively, each satisfying the volume-preserving condition, $\det\+F_{0,\ecm}=1$ and $\det\+F_{0,\fibre}=1$, and define $\+F_{\ecm}=\+F\+F_{0,\ecm}^{-1}$ and $\+F_{\fibre}=\+F\+F_{0,\fibre}^{-1}$. Correspondingly, $\log\+U_{\ecm}$ and $\*\l_\ecm$ will denote the elastic strain tensor---and its adapted Voigt representation---of the matrix component, and $\log\+U_{\fibre}$ and $\*\l_\fibre$ the elastic strain tensor---and its adapted Voigt representation---of the fibre component.

The kinematics of matrix and fibre components differ in the prescription of the (evolution of the) elastically-relaxed state. We take only one component of the two, the fibre, to be controlled by activation. In particular, we assume that the reference configuration is a relaxed one for the matrix, namely we will set $\+F_{0,\ecm}\equiv\mathbb{I}$, and that the field $\+F_{0,\fibre}$ depends on a control parameter $a$ which dictates whether the muscle is electrically stimulated or not. Since we are interested in capturing the passive state and the state of fused tetanic activation, we take $a$ to be a binary parameter, with $a=0$ and $a=1$ denoting the passive and activated states, respectively. In the absence of activation, $\+F_{0,\fibre}$ is the identity tensor, whereas in the presence of active contraction, it needs to be prescribed constitutively based on the selected experimental data.

The elastic response of the muscle tissue---viewed as a mixture of two materials---depends on the elastic strain of both the matrix and fibre components. We assume that the two contributions are independent and that the fibre component contributes to the mechanical response only in the presence of activation.
Specifically, we propose that the measured response results from the contribution of the matrix $\+S_\ecm=\hat{\+S}_\ecm(\log\+U_{\ecm})$ and, when active, of the fibre
$\+S_\fibre=\hat{\+S}_\fibre(\log\+U_{\fibre})$, and adopt an additive splitting of the total stress. This leads us to postulate the following model:
\[
\hat{\+S}_\tot(\log\+U_\ecm,\log\+U_\fibre)=\hat{\+S}_\ecm(\log\+U_{\ecm})+\hat{\+S}_\fibre(\log\+U_{\fibre}),
\]
such that $\hat{\+S}_\fibre\equiv0$ for $a=0$, and 
with the constraints of incompressibility and transverse isotropy on each component, which read
$\tr(\log\+U_\ecm)=0$ and $\tr(\log\+U_\fibre)=0$, and $\hat{\+S}_\ecm(\+Q^{\+T}\log\+U_{\ecm}\+Q)=\+Q^{\+T}\hat{\+S}_\ecm(\log\+U_{\ecm})\+Q$ and $\hat{\+S}_\fibre(\+Q^{\+T}\log\+U_{\fibre}\+Q)=\+Q^{\+T}\hat{\+S}_\fibre(\log\+U_{\fibre})\+Q$ for any rotation $\+Q$ about $\*l_1$.
In adapted coordinates, this model corresponds to an additive decomposition of the stress vector field $\hat{\*\sigma}_\tot:\RR^5\times\RR^5\to\RR^5$
\beq{eq:model_stress_matrix_fibre_decomposition}
  \hat{\*\sigma}_\tot(\*\l_\ecm,\*\l_\fibre)=\hat{\*\sigma}_\ecm(\*\l_\ecm)+\hat{\*\sigma}_\fibre(\*\l_\fibre),
\eeq
with $\hat{\*\sigma}_\ecm$ and $\hat{\*\sigma}_\fibre$ of the form~\eqref{eq:stress_incompressible_transverse}, and with $\hat{\*\sigma}_\fibre\equiv0$ for $a=0$. The prescription, from a data-based approach, of the overall ten material functions depending on eight invariants and of the activation term is the objective of the following Sections~\ref{sec:passive_model} and~\ref{sec:active_model}.

\subsection{Limitations and generalisability}
The strategy we propose for the constitutive prescription consists in identifying suitable material functions from measurements and, as such, the resulting model inherits the kinematic and loading-path restrictions of the data set it was built from. In particular, the constitutive assumptions introduced throughout the formulation are tailored to the experimental conditions considered, with the aim of reproducing the measured passive and active responses while retaining a mathematically consistent three-dimensional representation.

The assumption of vanishing hydrostatic pressure holds only under the specific kinematic and boundary conditions of the reference experiments---namely, quasi-static, homogeneous, uniaxial deformations of an unconstrained homogeneous specimen with no body forces. In a general three-dimensional boundary-value problem, the pressure---which, being a Lagrange multiplier, can not be constitutively prescribed---must be determined from the equilibrium equations and boundary conditions. In the presence of constraints or heterogeneous deformations, $p$ appears as an independent field in a more general version of equation~\eqref{eq:measured_stress} for the relation between Cauchy stress and second Piola–Kirchhoff components. Even so, the subsequent procedure for prescribing the constitutive stress components from data remains applicable once the hydrostatic pressure field is determined.

The assumption of additive splitting of the total stress into independent matrix and fibre contributions is also inherently due to experimental practice. The constitutive stress--strain relation for the fibre component $\hat{\*\sigma}_\fibre$ is identified directly from the data of~\cite{hawkins1994comprehensive}, computed as the difference between the stresses measured on stimulated and unstimulated tissue at a given stretch---a standard procedure for estimating the force generated by the contractile fibres in muscle tissue. While this procedure is supported by experimental evidence in uniaxial tests~\citep{winter2011whole}, the assumption it entails is known to affect both the data interpretation and the resulting model structure~\citep{rode2009effects}. More complex loading paths, such as shear deformations, might allow the modelling hypothesis of additive decomposition~\eqref{eq:model_stress_matrix_fibre_decomposition} to be tested and provide information on any mechanical coupling terms it neglects \citep{purslow2020structure}, for instance of the form $\hat{\*\sigma}_\fibre(\*\l_\fibre,\*\l_\ecm)$.
Even though in this work we do not need such terms, our procedure could be readily extended to treat similar cases.

We consider an activation controlled by a binary parameter $a\in\{0,1\}$. This choice limits the model to fully passive or fused tetanic active states, with no intermediate levels of contraction. This is consistent with the experimental data set considered here, which reports measurements of the maximal steady-state isometric force on fused tetanised tissue. To model more general physiological conditions such as sub-maximal stimulation or fatigue, and account for the dynamics of isometric contractions would require a time-dependent continuous activation parameter with an appropriate evolution equation (see e.g. \cite{giantesio2018loss,devita2017mathematical}).

Finally, we remark that the model's strain range follows directly from the data sets used. The passive data set covers tensile deformations up to $\l=0.25$, while the stress--strain relation in the activated case is identified from measurements covering Hencky strains spanning $\l\in[-0.4,0.4]$, with passive compressive stress set identically to zero, reflecting the tension--compression asymmetry of passive muscle tissue~\citep{bol2022multiscale}.
The model should therefore be regarded as validated only within this range---consistent with the physiological range of sarcomere length variation, and extrapolation beyond it is not supported by the data.

\section{Passive response}
\label{sec:passive_model}
The passive behaviour of skeletal muscle tissue results from its composite structure, hierarchically arranged in bundles of muscle fibres surrounded by multiple extracellular matrix (ECM) networks. At the tissue level, skeletal muscle appears to be transversely isotropic and incompressible, due to the high content of water \citep{takaza2013anisotropic}. Experiments performed on single fibres, ECM, or intact tissue highlight that all the components are closely interwoven, ultimately leading to an elastic response more complex than the sum of the various contributions. In fact, a nonlinear scaling of the measured stiffness at different scales has been reported: fascicles are stiffer than bundles that in turn are stiffer than fibres, suggesting an increasing relative importance of the ECM with scale \citep{ward2020scaling,lieber2021passive}.
Moreover, while the passive stretch of isolated fibre increases linearly with the strain, when longitudinally elongated, fibre bundles show a highly nonlinear response ascribable to the progressive engagement of the ECM \citep{meyer2011extracellular,purslow2020structure,ward2020scaling}. In particular, the response of the extracellular matrix alone appears nonlinear and more compliant in the along-fibre direction \citep{kohn2021direction}, in agreement with the mechanism of collagen recruitment common to many fibrous soft tissues. While many microstructural and multiscale models have been proposed to capture the complex mechanical response of the tissue from characterisations of the collagen network \citep{gindre2013structural,valentin2020inverse,miller2021microstructurally}, we shall use these experimental observations to explain our biophysically-informed macroscale model.

We construct our model for the passive elastic response of skeletal muscles upon the experimental data sets presented by \cite{takaza2013anisotropic}. 
Specifically, we rely on the data reported in Figure~\ref{fig:data}(a), which account for the dependence of the stress--strain relation on the fibre orientation. We exploit the theoretical setting presented in Section~\ref{sec:setting} to infer the dependence of the material functions on the system's invariants and propose a biophysically-based model for the passive quasi-static response of muscle tissue.

\subsection{Material functions}
In the passive regime, we set the control parameter to zero, i.e.\ $a=0$. Only the matrix component contributes to the mechanical response, since $\hat{\*\sigma}_\fibre$ vanishes by construction.
We start by prescribing the elastically-relaxed component of the deformation gradient for the matrix, $\+F_{0,\ecm}$. As anticipated in Section~\ref{sbsec:two-material_model}, we assume that the matrix is stress-free in the reference configuration, as we want the matrix contribution to be purely passive elastic. Thus, we set $\+F_{0,\ecm}\equiv\mathbb{I}$. Therefore, the strain of the matrix coincides with the measured strain and, since $\+F$ is a pure shear, the matrix strain is represented by $\log\+U_{\ecm}=\diag{\l,-\l/2,-\l/2}$. Since we are considering elongations, $\l>0$.
By computing the components of the strain on the adapted basis $\l_{\ecm,k}=\log\+U_{\ecm}:\+Z_k$, we obtain the matrix strain vector $\*\l_\ecm$ as a function of the measured stretch $\l$ and fibre inclination $\a$. The five components are
\begin{align*}
  \l_{\ecm,1}(\l,\a)&=\tfrac14\sqrt{\tfrac32}\l(1+3\cos(2\a)),
  &\l_{\ecm,4}(\l,\a)&=-\tfrac{3}{2\sqrt{2}}\l\sin(2\a),\\
  \l_{\ecm,2}(\l,\a)&=\tfrac{3}{2\sqrt{2}}\l\sin^2\a,
  &\l_{\ecm,5}(\l,\a)&=0.\\
  \l_{\ecm,3}(\l,\a)&=0,&
\end{align*}
The measured variables $\l$ and $\a$ are related to the strain invariants $\hat{\rho}_{\ecm,1}(\*\l_\ecm)=\l_{\ecm,1}$ and $\hat{\rho}_{\ecm}(\*\l_\ecm)=:\rho_{\ecm}$ by
\[
  \l=\hat{\l}(\l_{\ecm,1},\rho_\ecm)=\sqrt{\frac23}\rho_\ecm, \qquad
  \a=\hat{\a}(\l_{\ecm,1},\rho_\ecm)=\frac12\arccos\left(\frac43\frac{\l_{\ecm,1}}{\rho_\ecm}-\frac13\right).
\]
In particular, $\alpha$ and $\l$ are themselves invariants of the considered deformation. 

We assume that the material functions of our matrix model $c_{\ecm,1},\dots,c_{\ecm,5}$ depend on the two invariants $\l_{\ecm,1}$ and $\rho_\ecm$ only.
Explicitly, for $i=1,\dots,5$, we take
\beq{eq:hyp_1}
      c_{\ecm,i}=\hat{c}_{\ecm,i}(\l_{\ecm,1},\rho_\ecm).
\eeq
For the considered deformation, this hypothesis is equivalent to the condition $c_{\ecm,i}=\hat{c}_{\ecm,i}(\l,\a)$.
We furthermore postulate that
\beq{eq:hyp_2}
      c_{\ecm,4}\equiv0 \quad\text{ and }\quad
      c_{\ecm,5}\equiv0.
\eeq
The passive component of the measured Cauchy stress~\eqref{eq:measured_stress} under hypotheses~\eqref{eq:hyp_1} and~\eqref{eq:hyp_2}, as a function of the strain $\l$ and fibre inclination $\a$ is then
\begin{multline}\label{eq:tau_xx_tensile}
  \tau_{xx}^{\a,\ecm
  }(\l)=\frac{3}{4}\frac{\l}{2e^{-2\l}+e^\l}\\\times\left[\frac{(1+3\cos(2\a))^2}{4}\hat{c}_{\ecm,1}(\l,\a)+3(\sin\a)^4 \hat{c}_{\ecm,2}(\l,\a)+3(\sin(2\a))^2\hat{c}_{\ecm,3}(\l,\a)\right].
\end{multline}
In the following subsection, we prescribe the expressions for the material functions $\hat{c}_{\ecm,1}(\l,\a)$, $\hat{c}_{\ecm,2}(\l,\a)$, $\hat{c}_{\ecm,3}(\l,\a)$, deducing them from experimental data, biophysical arguments, and material symmetry conditions. Note that, by the combination of the representation~\eqref{eq:stress_incompressible_transverse} and hypothesis~\eqref{eq:hyp_2},  $c_{\ecm,1}$ and $c_{\ecm,3}$ determine the along-fibre normal and shear stresses, respectively, and $c_{\ecm,2}$ the cross-fibre normal and shear stresses.

\subsection{Constitutive model}
\begin{figure}[htb]
\centering
  \includegraphics[width=0.6\textwidth]{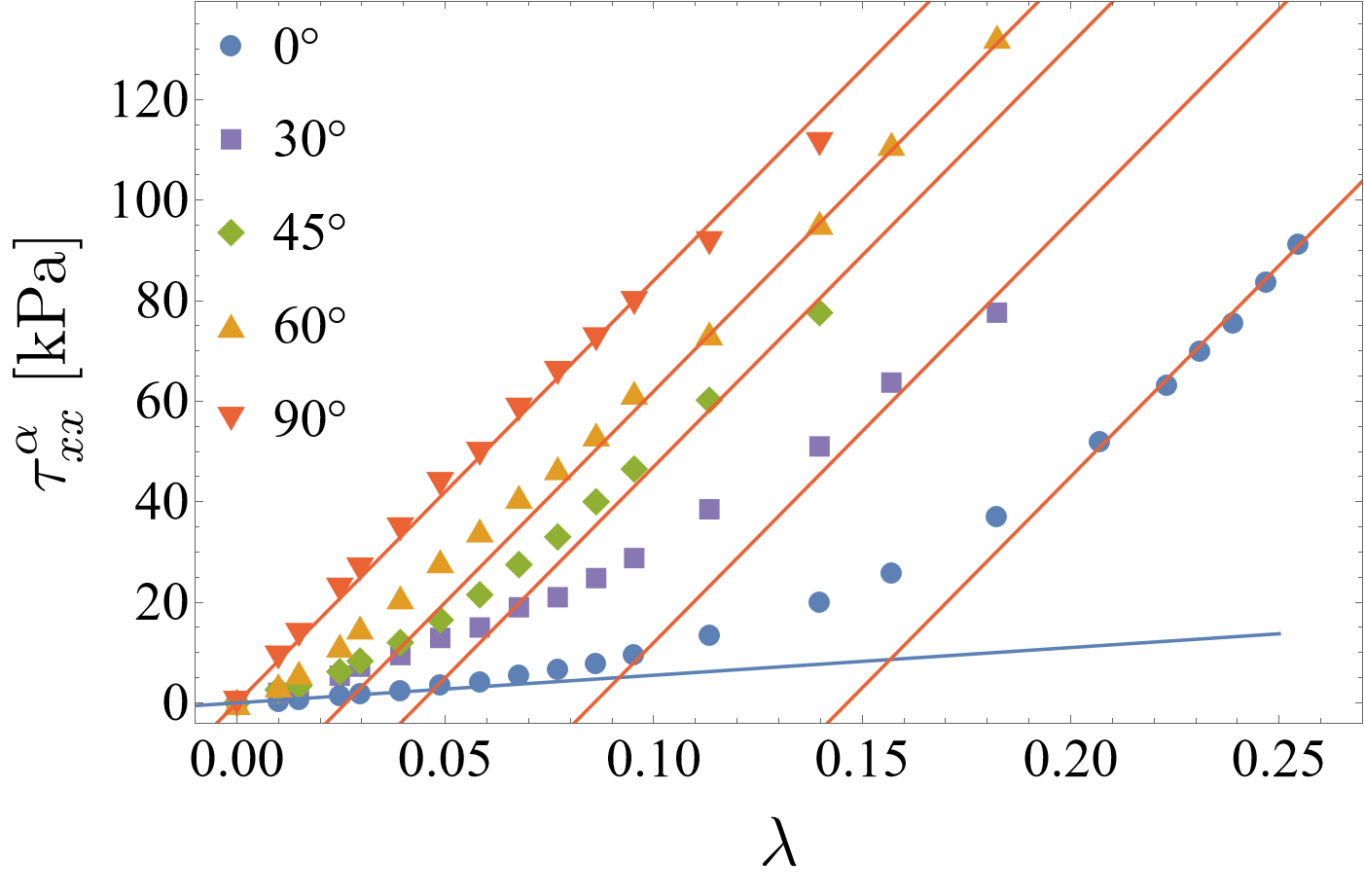}
\caption{Schematic representation of key features of the passive response, emerging from the experimental data set of~\cite{takaza2013anisotropic}, obtained in uniaxial tensile tests at different fibre alignments. The along-fibre stress increases nonlinearly with the strain (blue dots), displaying a transition between two linear regimes (blue and red lines). The measurements at inclination angles of $30^\circ,45^\circ,60^\circ$ show an analogous trend (purple, green, and orange dots) with however different stiffnesses at small strains. In the cross-fibre direction, the stress increases linearly with the strain (red dots). At large enough strains, the stiffness of the tissue appears to be the same regardless of the direction of stretching with respect to the fibres and coinciding with the cross-fibre one (red lines).}
    \label{fig:regions_passive}
\end{figure}

We observe the following features emerging from the data set of Figure~\ref{fig:data}(a) and reported in Figure~\ref{fig:regions_passive}.
For small strains, the elastic response appears linear, with a stiffness that depends on the orientation of the fibres with respect to the direction of stretching. In particular, the longitudinal direction is considerably more compliant than the transverse one, and at intermediate inclinations the stress--strain curves fall in between these two. The linear trend persists even at larger strains for stretching orthogonal to the fibre direction, while the behaviour at every other angle is nonlinear. However, all data show a linear growth for large enough strain values---dependent on the fibre orientation---with a common slope which coincides with the stiffness measured in the cross-fibre direction. This linear regime is attained later for smaller fibre inclination.
These observations lead us to propose the following fitting function for the stress--strain relation:
\begin{subequations}\label{eq:tau_xx_fit_tensile}
\beq{eq:tau_xx_fit_tensile_1}
      \tau_{\fit}^\ecm(\l,\a)=k^\ecm(\l,\a)\l
\eeq
with
\beq{eq:tau_xx_fit_tensile_2}
      k^\ecm(\l,\a)=m^\ecm_l(\a)+\frac{r^\ecm(\a)}{s^\ecm(\a)}\frac{m^\ecm_t-m^\ecm_l(\a)}{\l}\log\left(\frac{e^{s^\ecm(\a)}+e^{\tfrac{s^\ecm(\a)}{r^\ecm(\a)}\l}}{e^{s^\ecm(\a)}+1}\right),
\eeq
\end{subequations}
where $m^\ecm_l,m^\ecm_t,r^\ecm,s^\ecm$ are real parameters. 
The splitting into two terms reflects the different regimes, characterised by different stiffness, discussed above. Indeed, the first term accounts for the linear response in $\l$ at small strains, with slope depending on $\a$, as $\lim_{\l\to0}\partial_\l\tau_{\fit}^\ecm(\l,\a)=m^\ecm_l(\a)$. On the other hand, the second term dominates at larger strains, in particular, $\lim_{\l\to+\infty}k^\ecm(\l,\a)=m^\ecm_t$ for any fibre orientation $\a$. To account for a smooth transition between the two linear growths in the strain, with constant slopes $m^\ecm_l$ and $m^\ecm_t$ respectively, we integrated the sigmoidal logistic function $\delta(\l):=m^\ecm_l+\frac{m^\ecm_t-m^\ecm_l}{1+e^{-s^\ecm(\l-r^\ecm)/r^\ecm}}$. The parameter $r^\ecm$ is the function's midpoint, at which the steepness is maximal, and $s^\ecm/r^\ecm$ gives the growth rate.
We remark that, by viewing the stress as a function of both the strain and the fibre orientation, we are able to fit the data sets with just one interpolating function.
    
\begin{figure}[htb]
\centering
\includegraphics[width=0.7\textwidth]{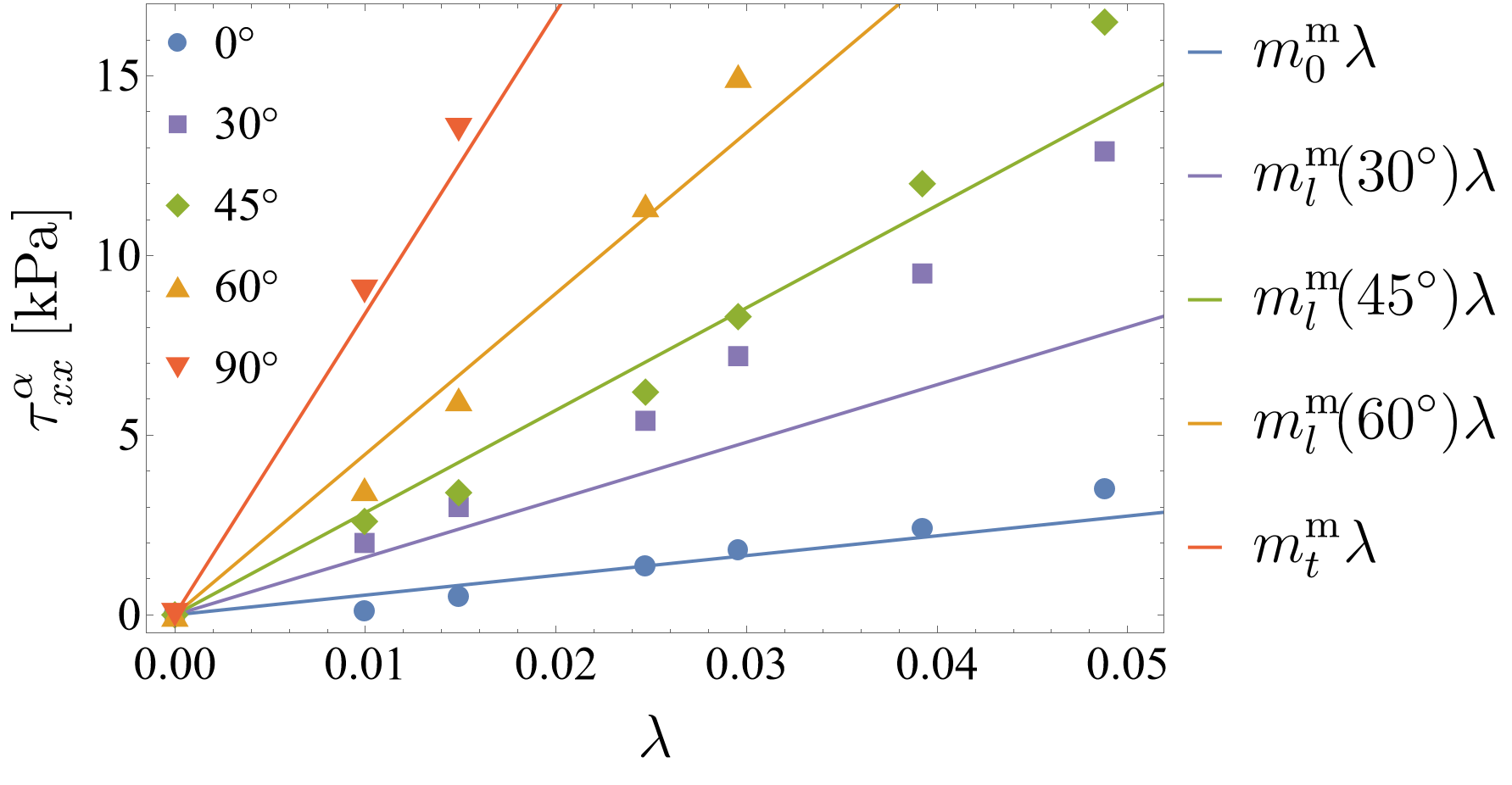}
\caption{Fit of the experimental data in the passive regime at small strains. The proposed stress--strain fitting function~\eqref{eq:tau_xx_fit_tensile} is linear in the strain, for sufficiently small $\l$, with stiffness $m_l^\ecm(\a)$ dependent on the orientation of the fibre~\eqref{eq:mu_f_alpha_tensile}. The two material parameters $m^\ecm_0,m^\ecm_t$ are the longitudinal and transverse stiffness, respectively, and are obtained by interpolation of the corresponding data (blue and red lines, respectively). The remaining (purple, green, and orange) curves are, instead, predictions of the model.}
\label{fig:fit_tensile_small_strains}
\end{figure}

To construct a biophysically-based model, we relate the mechanical experimental evidence and the introduced fitting to our knowledge of the microstructure of the tissue. 
The independence of the stiffness of the tissue from the fibre orientation at large strains indicates that, in that regime, the extracellular matrix is likely the main contributor to the mechanical response. 
However, the relative contribution of the two components of the tissue is relevant at small strains. 
In this regime, we argue that the fibres do not resist stretching and all the response is due to the matrix. The increase in stiffness with the inclination of the fibres can then be explained in terms of the effective area of the element being stretched that pertains to the matrix. 
Assuming that the ratio between the effective matrix area for along-fibre deformations and the total area equals the ratio between the longitudinal and transverse stiffnesses, namely $A_{\text{eff}}(0)/A_{\text{tot}}=m^\ecm_0/m^\ecm_t$, then the effective area when the fibres are inclined by an angle $\a$ with respect to the direction of stretching is given by $A_{\text{eff}}(\a)=A_{\text{tot}}-(A_{\text{tot}}-A_{\text{eff}}(0))\cos\a$.
This reasoning leads us to propose the following dependence on the orientation of the fibres of the fitting coefficient $m^\ecm_l$ in~\eqref{eq:tau_xx_fit_tensile_2} in the linear elastic regime at small strains:
\beq{eq:mu_f_alpha_tensile}
  m^\ecm_l(\a)=m^\ecm_t\left(1-\left(1-\frac{m^\ecm_0}{m^\ecm_t}\right)\cos\a\right).
\eeq
From a linear fitting of the data in the along- and cross-fibre directions, at small strains, we obtain the interpolating coefficients $m^\ecm_0$ and $m^\ecm_t$. Their values, together with their statistical error, obtained from the linear regression are reported in Table~\ref{tab:par_passive}. We then test relation~\eqref{eq:mu_f_alpha_tensile} on the remaining data sets: the result is presented in Figure~\ref{fig:fit_tensile_small_strains}. The good agreement of the experimental data with the fitting function $\tau_{\fit}^\ecm(\l,\a)$, for any fibre orientation, at small $\l$, validates the mechanical mechanism that we propose, as it predicts accurately the dependence $m^\ecm_l(\a)$ of the stiffness on $\a$.
        
The nonlinear transition between the two linear regimes is described by the coefficients $r^\ecm(\a)$ and $s^\ecm(\a)$, which, by construction, quantify the characteristic strain at which the transition occurs and its width, and depend on the fibre orientation. We interpolate the data at the various inclination angles with the fitting function~\eqref{eq:tau_xx_fit_tensile} to extrapolate the dependence of these coefficients on $\a$.
\begin{figure}[htb]
\centering
\includegraphics[width=0.508\textwidth]{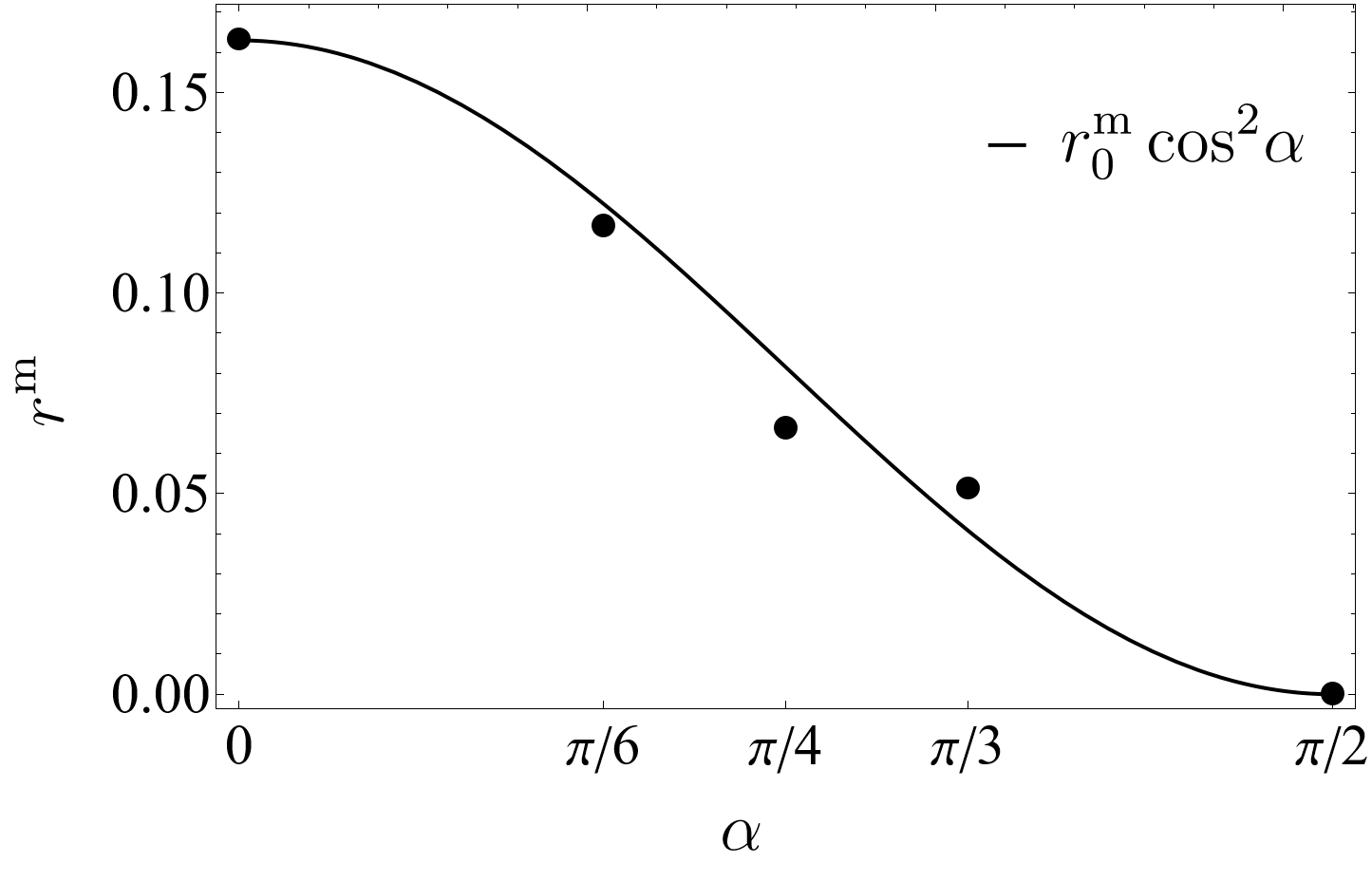}
\includegraphics[width=0.478\textwidth]{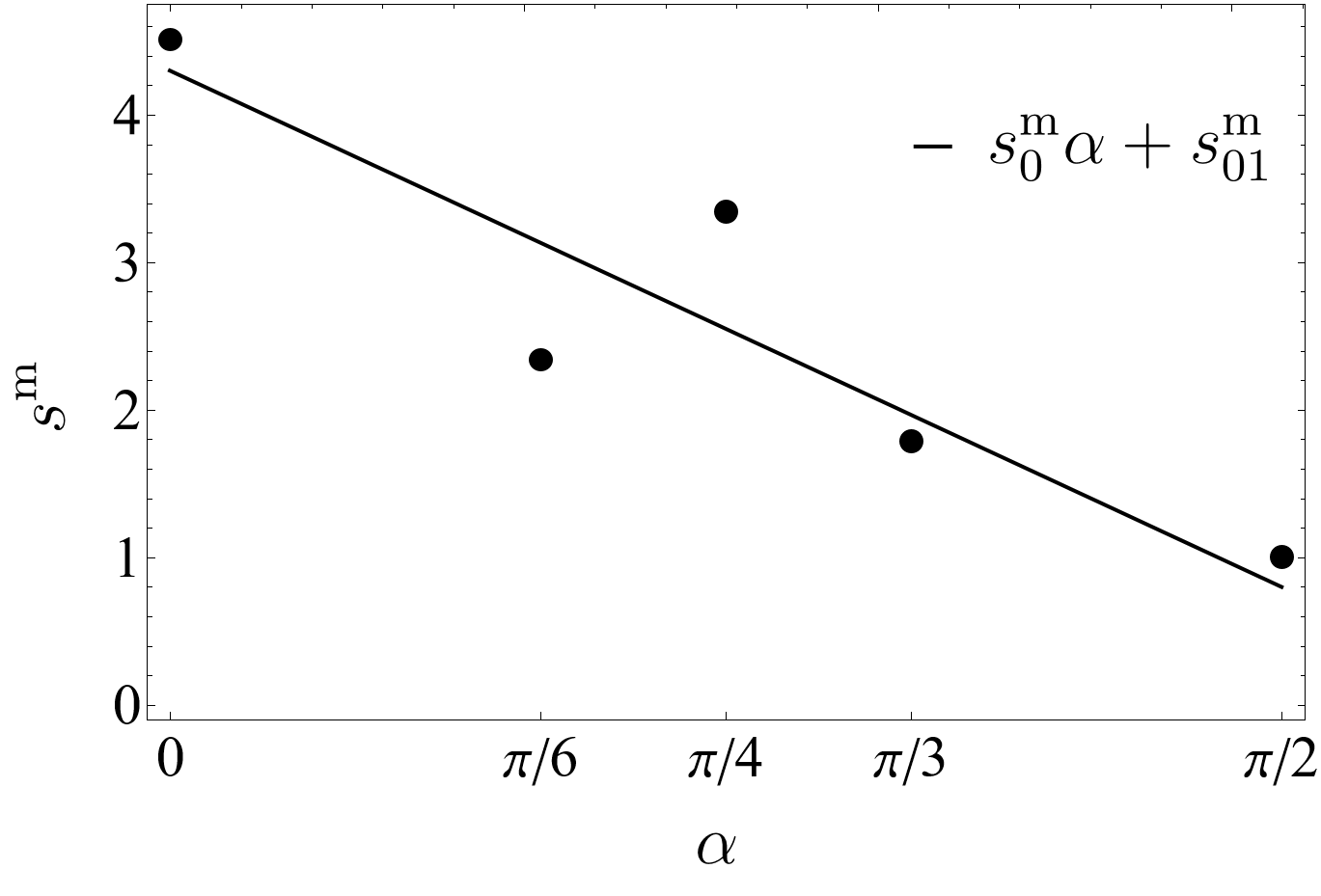}
\caption{The coefficients $r^\ecm$ and $s^\ecm$, which account for the nonlinear transition between the two linear regimes in~\eqref{eq:tau_xx_fit_tensile}, as functions of the fibre orientation. The values of $r^\ecm$ and $s^\ecm$ (black dots) are interpolating parameters from fitting~\eqref{eq:tau_xx_fit_tensile} against the data. Solid curves represent interpolating functions of such parameters depending on $\alpha$.}
\label{fig:fit_tensile_a_b_alpha}
\end{figure}
Figure~\ref{fig:fit_tensile_a_b_alpha} shows that $r^\ecm$ is a decreasing function of $\a$, and the data are well interpolated by $r^\ecm(\a)=r^\ecm_0(\cos\a)^2$. We interpret this trend by conceiving of the collagen network as engaging in the elastic response at larger strains when the tissue is stretched in the direction of the fibres, with an increasing portion of the system being progressively recruited by the deformation, while the extracellular matrix contributes immediately to the response when stretched orthogonally to the fibres.
As for $s^\ecm$, the values obtained are not significant enough, and we interpolate the data with a linear regression, $s^\ecm(\a)=s^\ecm_0\a+s^\ecm_{01}$.
Ultimately, the passive part of the model is fully characterised by five material parameters, to be determined through the fitting function~\eqref{eq:tau_xx_fit_tensile} with the experimental data set: $m^\ecm_0$, $m^\ecm_t$, $r^\ecm_0$, $s^\ecm_0$, $s^\ecm_{01}$. Table~\ref{tab:par_passive} collects the values of such parameters obtained through the fitting procedure on the reference experimental data by~\cite{takaza2013anisotropic}.

\begin{table}[htb]
\caption{Values of the model parameters entering the fitting function~\eqref{eq:tau_xx_fit_tensile} for the experimental data set of~\cite{takaza2013anisotropic}. The parameters $m^\ecm_0$ and $m^\ecm_t$ are obtained from linear regressions in $\l$ (Figure~\ref{fig:fit_tensile_small_strains}), $r^\ecm_0$ in $\cos^2\a$, and $s^\ecm_0$ and $s^\ecm_{01}$ in $\a$ (Figure~\ref{fig:fit_tensile_a_b_alpha}).}
\centering
\begin{tabular}{c c c} 
 \toprule
 Parameter & Value & Standard error\\ [0.5ex] 
 \midrule
 $m^\ecm_0$ [kPa] & 56 & 5\\
 $m^\ecm_t$ [kPa] & 840 & 8\\
 $r^\ecm_0$ & 0.158 & 0.007\\
 $s^\ecm_0$ & -2.1 & 0.6\\
 $s^\ecm_{01}$ & 4.3 & 0.5\\[0.5ex]
 \bottomrule
\end{tabular}
\label{tab:par_passive}
\end{table}

We next exploit the experimental data of the measurements with fibre orientation at $0^\circ,45^\circ,90^\circ$ to infer the functional dependence of the material functions on the invariants, from~\eqref{eq:tau_xx_tensile} and~\eqref{eq:tau_xx_fit_tensile}.

\begin{figure}[htb]
\centering
\includegraphics[width=0.8\textwidth]{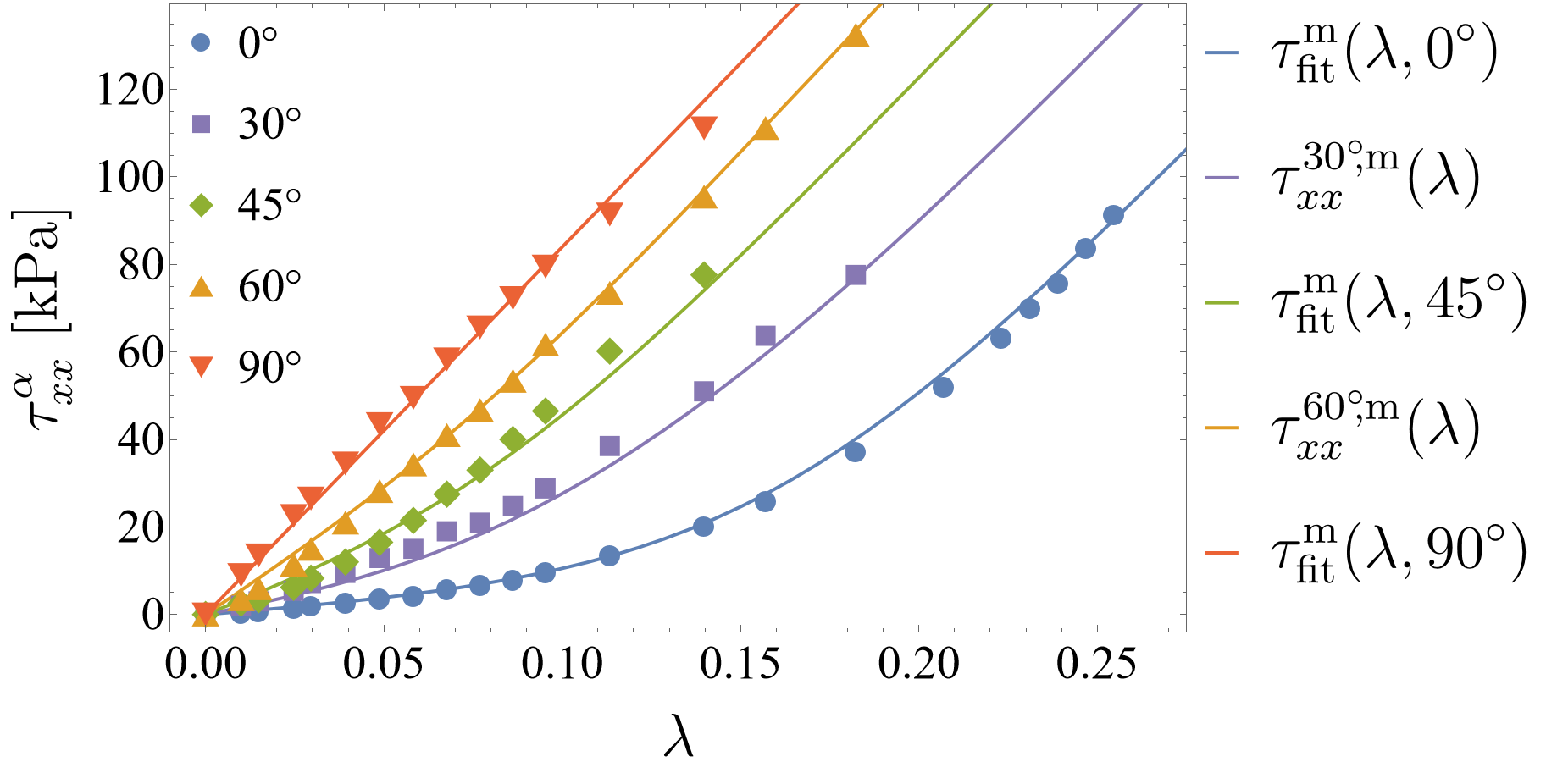}
\caption{Fitting of the model~\eqref{eq:model_tensile} on the data by~\cite{takaza2013anisotropic} for the passive response of the muscle tissue. The measurements associated with stretching at fibre orientations of $0^\circ,45^\circ,90^\circ$ have been used to identify the material functions~\eqref{eq:material_functions_c_passive} and calibrate the material parameters in~\eqref{eq:tau_xx_fit_tensile}, while we test the predictive ability of the proposed model on the measurements at $30^\circ$ and $60^\circ$.}
\label{fig:fit_tensile}
\end{figure}

\begin{itemize}
\item With $\a=0$, from $\tau_{xx}^{0,\ecm}(\l)=\tau_{\fit}^\ecm(\l,0)$, we get
\[
    \hat{c}_{\ecm,1}(\l,0)=\frac{2e^{-2\l}+e^\l}{3}k^\ecm(\l,0),
\]
and we make the hypothesis
\beq{eq:hyp_3}
     \hat{c}_{\ecm,1}(\l,\a)=\hat{c}_{\ecm,1}(\l,0)\equiv \hat{c}_{\ecm,1}(\l).
\eeq
    
\item With $\a=\pi/2$, from $\tau_{xx}^{\pi/2,\ecm}(\l)=\tau_{\fit}^\ecm(\l,\pi/2)$, we get
\[
    \hat{c}_{\ecm,2}(\l,\pi/2)=\frac{4(2e^{-2\l}+e^\l)}{9}k^\ecm(\l,\pi/2)-\frac13 \hat{c}_{\ecm,1}(\l),
\]
and we make the hypothesis
\beq{eq:hyp_4}
    \hat{c}_{\ecm,2}(\l,\a)=\hat{c}_{\ecm,2}(\l,\pi/2)\equiv \hat{c}_{\ecm,2}(\l).
\eeq
    
\item With $\a=\pi/4$, from $\tau_{xx}^{\pi/4,\ecm}(\l)=\tau_{\fit}^\ecm(\l,\pi/4)$, we get
\[
    \hat{c}_{\ecm,3}(\l,\pi/4)=\frac{4(2e^{-2\l}+e^\l)}{9}k^\ecm(\l,\pi/4)-\frac{1}{12} \hat{c}_{\ecm,1}(\l)-\frac14 \hat{c}_{\ecm,2}(\l),
\]
and we make the hypothesis
\beq{eq:hyp_5}
    \hat{c}_{\ecm,3}(\l,\a)=\hat{c}_{\ecm,3}(\l,\pi/4)\equiv \hat{c}_{\ecm,3}(\l).
\eeq
\end{itemize}
In terms of the invariants $\l_{\ecm,1}$ and $\rho_\ecm$, we thus obtain material functions that only depend on $\rho_\ecm$ and have the following expressions:
\beq{eq:material_functions_c_passive}
\begin{gathered}
    \hat{c}_{\ecm,1}(\rho_\ecm)=\frac{2e^{-2\sqrt{\frac23}\rho_\ecm}+e^{\sqrt{\frac23}\rho_\ecm}}{3}k^\ecm\Bigl(\sqrt{\tfrac23}\rho_\ecm,0\Bigr),
\\
    \hat{c}_{\ecm,2}(\rho_\ecm)=\frac{2e^{-2\sqrt{\frac23}\rho_\ecm}+e^{\sqrt{\frac23}\rho_\ecm}}{9}\left(4k^\ecm\Bigl(\sqrt{\tfrac23}\rho_\ecm,\frac\pi2\Bigr)-k^\ecm\Bigl(\sqrt{\tfrac23}\rho_\ecm,0\Bigr)\right),
\\
    \hat{c}_{\ecm,3}(\rho_\ecm)=\frac{2e^{-2\sqrt{\frac23}\rho_\ecm}+e^{\sqrt{\frac23}\rho_\ecm}}{9}\left(4k^\ecm\Bigl(\sqrt{\tfrac23}\rho_\ecm,\frac\pi4\Bigr)-k^\ecm\Bigl(\sqrt{\tfrac23}\rho_\ecm,\frac\pi2\Bigr)\right).
\end{gathered}
\eeq
Ultimately, from~\eqref{eq:stress_incompressible_transverse},~\eqref{eq:hyp_2}, and~\eqref{eq:material_functions_c_passive}, we arrive at the following constitutive law for the matrix component:
\beq{eq:model_tensile}
    \hat{\*\sigma}_\ecm(\*\l_\ecm)=
    \hat{c}_{\ecm,1}(\hat{\rho}_\ecm(\*\l_\ecm))\!\mx{\l_{\ecm,1}\\0\\0\\0\\0}
    +\hat{c}_{\ecm,2}(\hat{\rho}_\ecm(\*\l_\ecm))\!\mx{0\\\l_{\ecm,2}\\0\\0\\\l_{\ecm,5}}
    +\hat{c}_{\ecm,3}(\hat{\rho}_\ecm(\*\l_\ecm))\!\mx{0\\0\\\l_{\ecm,3}\\\l_{\ecm,4}\\0}.
\eeq
Figure~\ref{fig:fit_tensile} shows the final fits with the function $\tau_{\fit}^\ecm(\l,\alpha)$, defined by~\eqref{eq:tau_xx_fit_tensile}, of the measured stress corresponding to stretches at $\a=0^\circ,45^\circ,90^\circ$ with respect to the fibre orientation (blue, green, and red curves), with material parameters given in Table~\ref{tab:par_passive}.
The two remaining data sets, collecting stress--strain measurements at intermediate inclinations, allow us to test our model~\eqref{eq:model_tensile}. The purple and orange curves represent the function $\tau_{xx}^{\a,\ecm}(\l)$ given in~\eqref{eq:tau_xx_tensile} for $\a=30^\circ$ and $\a=60^\circ$, respectively, with material functions given by~\eqref{eq:material_functions_c_passive}.
The excellent agreement of the predicted stress-strain relation with the experimental data for these intermediate inclinations provides a solid validation of the proposed model for the passive response of the tissue.

\begin{figure}[htb]
\centering
\includegraphics[width=\textwidth]{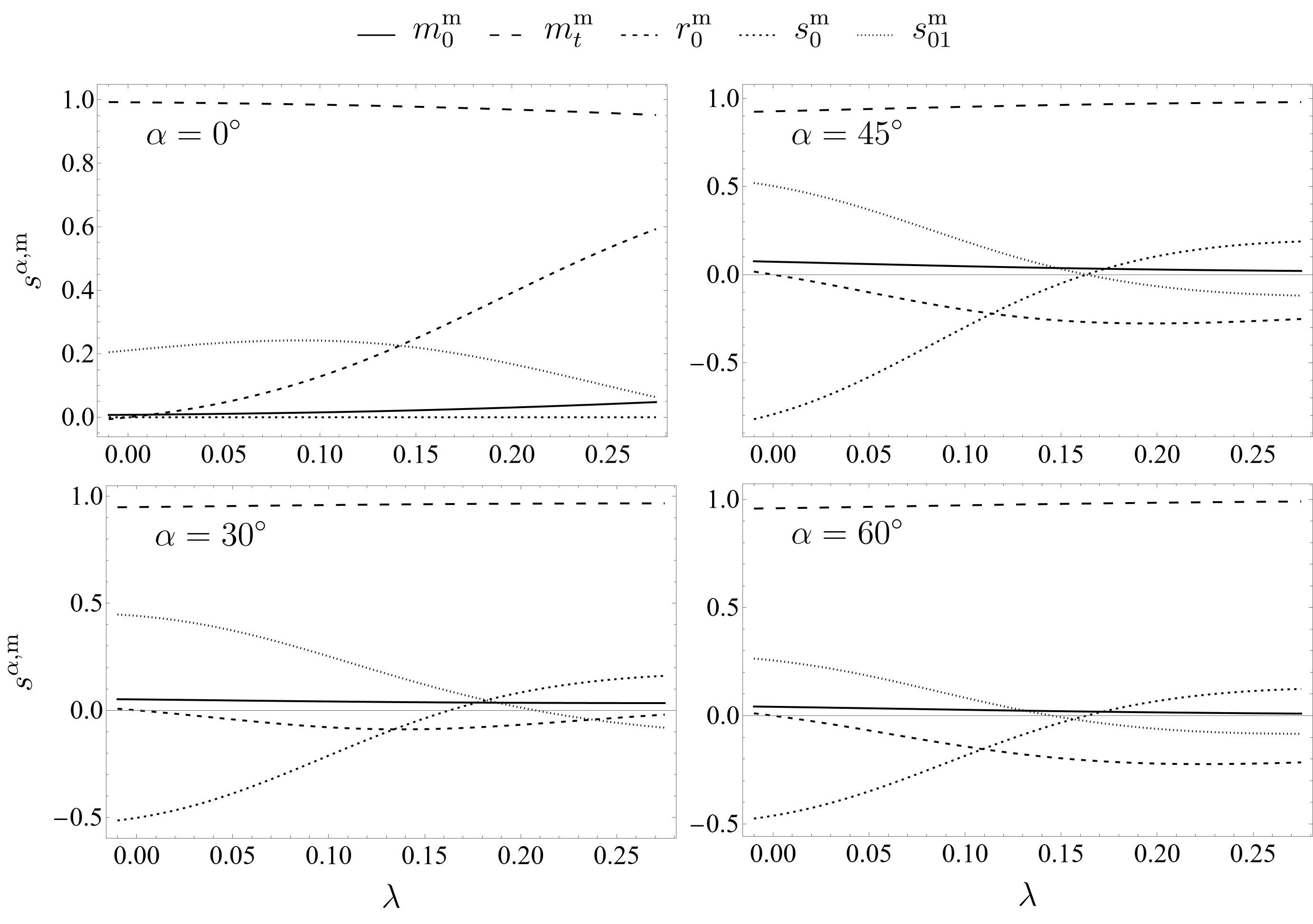}
\caption{Normalised local sensitivity coefficient $s^{\a,\ecm}(\l,q)$, which provides us an estimate of the relative contributions of the different model parameters $q=m^\ecm_0, m^\ecm_t, r^\ecm_0, s^\ecm_0, s^\ecm_{01}$ on the constitutive stress--strain relation~\eqref{eq:tau_xx_tensile}. While for $\a=90^\circ$ the passive response depends on the parameter $m^\ecm_t$ only, for the other angles $m^\ecm_t$ remains consistently the most influential parameter across the whole stretch range, while the other parameters contribute non-trivially, with their relative weight varying as a function $\l$.}
\label{fig:sens_tensile}
\end{figure}

\begin{figure}[htb]
\centering
\includegraphics[width=0.8\textwidth]{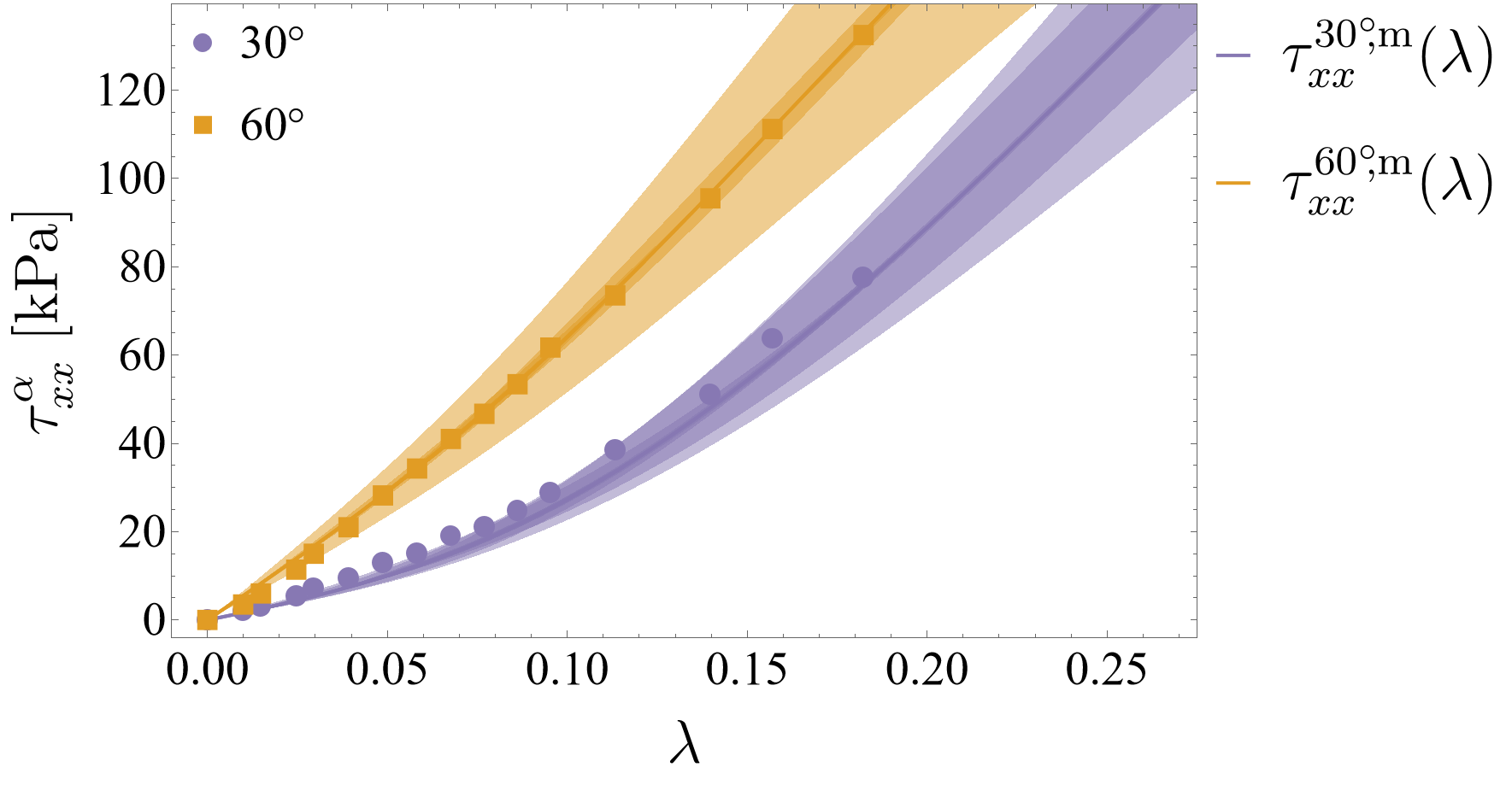}
\caption{Perturbed stress--strain curves~\eqref{eq:tau_xx_tensile} at $\a=30^\circ,60^\circ$, obtained by varying each single model parameter by $\pm20\%$ (shaded bands), to assess the influence of uncertainty in the material parameters. The bands for the different parameters are plotted as superimposed, semi-transparent regions, so darker areas reflect the overlap of multiple perturbations. The predictions of our model for the passive tensile response remain reasonably robust to parameter variations.}
\label{fig:sens_shaded_tensile}
\end{figure}

To estimate the relative weight of each material parameter $q=m^\ecm_0, m^\ecm_t, r^\ecm_0, s^\ecm_0, s^\ecm_{01}$ entering the stress--strain relation~\eqref{eq:tau_xx_fit_tensile}, with the prescribed material functions~\eqref{eq:material_functions_c_passive}, we compute the normalised local sensitivity coefficient $s^{\a,\ecm}(\l,q)=\frac{\partial \tau_{xx}^{\a,\ecm}(\l)}{\partial q}\cdot\frac{q}{\tau_{xx}^{\a,\ecm}(\l)}$, and plot it, as a function of the stretch $\l$, for each fibre inclination $\a$ (see Figure~\ref{fig:sens_tensile}). We can, in particular, notice that, at any angle, the model results more sensitive to variations of the parameter $m_t^\ecm$, representing the stiffness of the extracellular matrix in the cross-fibre direction. Figure~\ref{fig:sens_shaded_tensile} shows the perturbed stress--strain curves~\eqref{eq:tau_xx_tensile} at $\a=30^\circ,60^\circ$ obtained by varying the model parameters by $\pm20\%$, providing a quantification of the robustness of the model's predictions.

\section{Active response}
\label{sec:active_model}
Active muscle contraction, which generates internal stress even in the absence of external loading, results from the shortening of its muscle fibres, through the conversion of chemical energy into mechanical work. This mechanism is well explained by the sliding filament and cross-bridge theories, which describe how sarcomeres shorten as a consequence of ATP-driven cross-bridge cycling between parallel actin and myosin filaments, regulated by calcium ions~\citep{huxley1957muscle,mukund2020skeletal,powers2021sliding}. 
A gold standard benchmark contribution by \cite{gordon1966variation} experimentally established the steady-state active force–length relation of skeletal muscle and showed that its characteristic dependence on muscle length is explained by the degree of overlap between actin and myosin filaments, thereby providing support for the sliding filament theory. Since sarcomeres are arranged in series along each muscle fibre, their shortening accumulates, producing axial shortening of the entire fibre. Active force generation is therefore confined to the fibre direction, while transverse deformation arises passively as a consequence of the near-incompressibility of muscle tissue. These features provide the physiological basis for our choice of active stress contribution. While many multiphysics models have been developed to capture active force generation from sarcomere dynamics \citep{julian1969activation,chapelle2012energy,regazzoni2020biophysically}, our objective is instead to formulate a macroscale constitutive model that reproduces the experimentally observed steady-state response. To this end, we construct the model for the fibre component using the uniaxial along-fibre measurements of~\cite{hawkins1994comprehensive}, obtained under isometric tetanic conditions (Figure~\ref{fig:data}(b)).

\subsection{Active strain tensor}
When we switch on the control parameter, setting $a=1$, we model the response of a fully tetanised tissue, in contrast to the passive regime in which $a=0$, while no intermediate degrees of contraction are considered. We assume that the active contraction occurs parallel to the fibre direction, namely we prescribe the following elastically-relaxed component of the deformation gradient $\+F_{0,\fibre}$:
\[
  \+F_{0,\fibre}=\mx{e^{a\bar\l}&0&0\\0&e^{-a\bar\l/2}&0\\0&0&e^{-a\bar\l/2}},
\]
where the parameter $\bar{\l}$ characterises the stretch ratio at which the fibre component is relaxed. Indeed, the elastically-relaxed configuration of the fibre component is obtained by imposing $\+F=\+F_{0,\fibre}$, which gives the condition $\l=a\bar{\l}$.
With this choice, the fibre strain is represented by the diagonal matrix $\log\+U_{\fibre}=\diag{\l-a \bar\l,(a \bar\l-\l)/2,(a \bar\l-\l)/2}$. While we take the activation control parameter to be a constant, we shall assume that the elastically-relaxed strain is constitutively dependent on the deformation.
The dependence of $\+F_{0,\fibre}$ on $\+F$ will be postulated on the basis of the selected data.

\subsection{Material functions}
The available data set that investigates the active response consists of uniaxial deformations along the direction of the fibres, hence we set $\a=0$.
By computing the components of the fibre strain on the adapted basis $\l_{\fibre,k}=\log\+U_{\fibre}:\+Z_k$, $k=1,\dots,5$, we obtain the fibre strain vector $\*\l_\fibre$ as a function of the measured stretch $\l$. The only non-vanishing component is the along-fibre stretch: $\*\l_\fibre=(\l_{\fibre,1}(\l),0,0,0,0)$ with
\[
  \l_{\fibre,1}(\l)=\sqrt{\tfrac32}(\l-\bar{\l}).
\]
This implies that, for this deformation, $\hat{\rho}_{\fibre,1}(\*\l_\fibre)=\hat{\rho}_{\fibre}(\*\l_\fibre)=\l_{\fibre,1}(\l)$ (or, equivalently, $\l-\bar{\l}$) is the only invariant of the fibre strain and that the only component of the fibre stress that contributes to the elastic response is $\sigma_{\fibre,1}$. 
Thus, we postulate that the material functions for the active part of our model satisfy
\beq{eq:hyp_6}
      c_{\fibre,1}=\hat{c}_{\fibre,1}(\l_{\fibre,1}),
\eeq
which, for the considered deformation, is equivalent to $c_{\fibre,1}=\hat{c}_{\fibre,1}(\l-\bar{\l})$,
and
\beq{eq:hyp_7}
      c_{\fibre,i}\equiv0 \quad\text{ for }i=2,\dots,5.
\eeq
The active component of the measured Cauchy stress~\eqref{eq:measured_stress} under hypotheses~\eqref{eq:hyp_6} and~\eqref{eq:hyp_7}, as a function of the strain $\l$ is then
\beq{eq:tau_xx_fibre}
  \tau_{xx}^{0,\fibre}(\l)=\frac{3(\l-\bar{\l})}{2e^{-2\l}+e^\l}\hat{c}_{\fibre,1}(\l-\bar{\l}).
\eeq
Furthermore, we assume that $\bar{\l}$ is itself a function of $\l$.
In the following subsection, we prescribe the expression for the material function $\hat{c}_{\fibre,1}(\l-\bar{\l})$---that determines the along-fibre normal stress, and the dependence of the elastically-relaxed configuration of the fibre component $\bar{\l}$ on the stretch ratio $\l$.

As for the matrix component, which is also involved in the active response, from the model constructed in Section~\ref{sec:active_model}, we have $\*\l_\ecm=(\l_{\ecm,1}(\l,0),0,0,0,0)$ with $\l_{\ecm,1}(\l,0)=\sqrt{\tfrac32}\l$ and $c_{\ecm,1}$ given by~\eqref{eq:material_functions_c_passive}.
The dependence of the passive component of the Cauchy stress on the strain $\l$ is:
\[
  \tau_{xx}^{0,\ecm}(\l)=\frac{3\l}{2e^{-2\l}+e^\l}\hat{c}_{\ecm,1}(\l).
\]
The total stress is then obtained by adding the two contributions, owing to the constitutive hypothesis~\eqref{eq:model_stress_matrix_fibre_decomposition}, by which $\hat{\sigma}_{\tot,1}(\l)=\hat{\sigma}_{\ecm,1}(\l)+\hat{\sigma}_{\fibre,1}(\l)$. Namely,
\[
  \tau_{xx}^{0,\tot}(\l)=\frac{\sqrt{6}}{2e^{-2\l}+e^\l}\hat{\sigma}_{\tot,1}(\l)=\tau_{xx}^{0,\ecm}(\l)+\tau_{xx}^{0,\fibre}(\l).
\]

\subsection{Constitutive model}
From the data set reported in Figure~\ref{fig:data}(b), we shall now extract constitutive relations for $c_{\fibre,1}$ and $\bar{\l}$. In particular, we exploit the data for the active stress, which was estimated by~\cite{hawkins1994comprehensive} by subtracting from the (total) isometric stress at a certain length the passive stress measured without stimulation at the same length. As discussed by~\cite{rode2009effects}, this computation corresponds to assuming that the stress produced by the active contraction of the fibres superimposes to the passive response.

\begin{figure}[htb]
    \centering
    \includegraphics[width=0.6\textwidth]{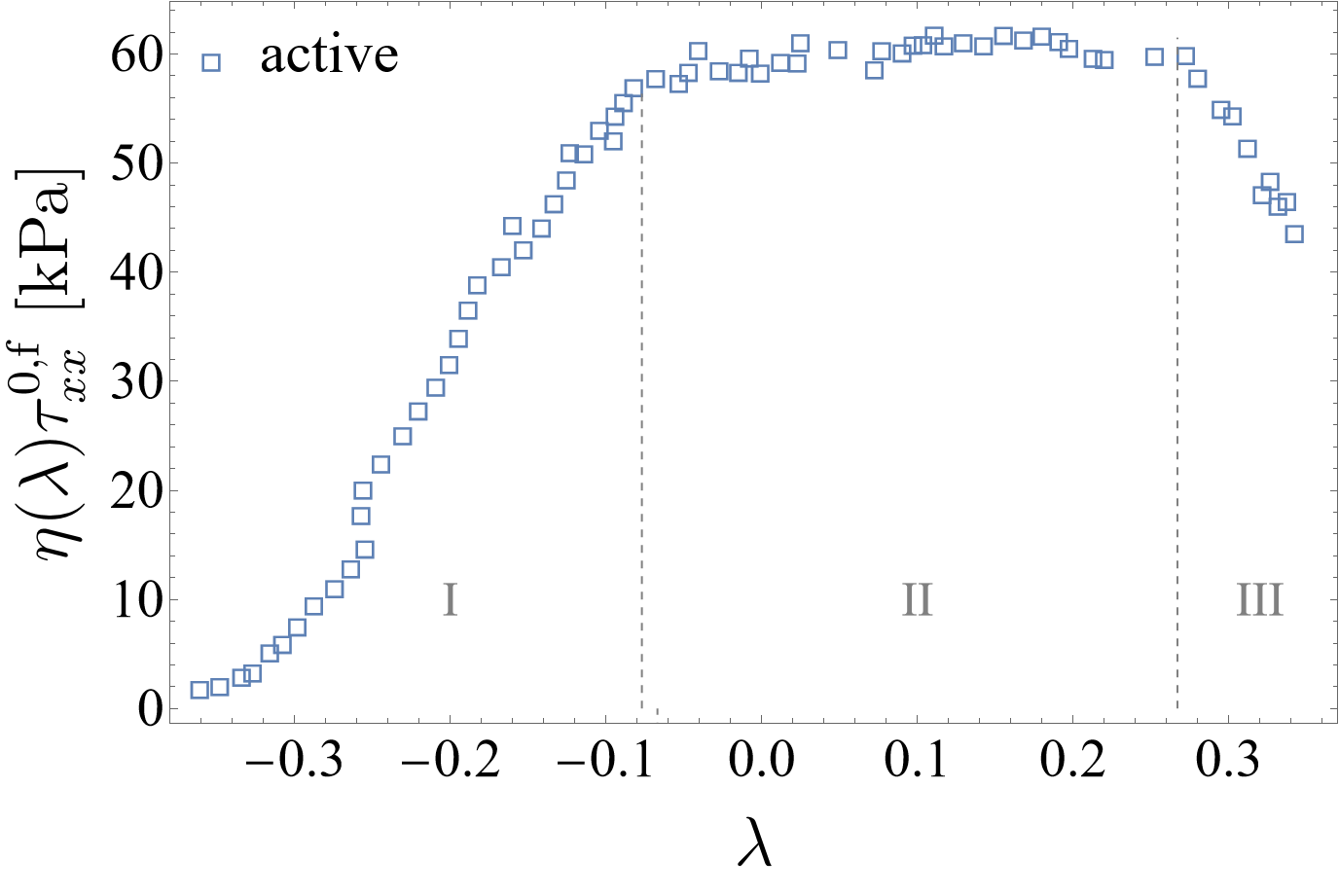}
    \caption{Schematic subdivision of different regimes emerging in the active response, identifiable in the experimental data set by~\cite{hawkins1994comprehensive}, obtained in uniaxial tensile tests in the fibre direction of tetanised samples. Each stress $\tau_{xx}^{0,\fibre}(\l)$ is rescaled by a factor $\eta(\l)$. At small strain with respect to the active elastically-relaxed configuration (region I), the response of the tissue is nonlinearly elastic, whereas the active material displays a typically plastic behaviour as the applied strain increases. Specifically, we can identify perfectly plastic (region II) and strain-softening (region III) phases. In both plastic regions, the stress-strain dependence appears linear, with different slopes.}
    \label{fig:regions_active}
\end{figure}

Differently from the passive response analysis, rather than identifying suitable fits for the pairs $(\l,\tau_{xx}^{0,\fibre}(\l))$, we now first rescale each measured stress $\tau_{xx}^{0,\fibre}(\l)$ by a factor $\eta(\l):=(2e^{-2\l}+e^{\l})/3$. Such an expedient facilitates the prescription of a fitting function satisfying the material symmetry: indeed, the material function $\hat{c}_{\fibre,1}$, which we want to prescribe, and which is related to the measured stress by equation~\eqref{eq:tau_xx_fibre}, can depend on the difference $\l-\bar{\l}$ only. We hence consider the rescaled data set of the pairs $\left(\l,\eta(\l)\tau_{xx}^{0,\fibre}(\l)\right)$ reported in Figure~\ref{fig:regions_active}.

We observe the following features, typical of muscle force-length relations \citep{rassier1999length}.
The (rescaled) active component of the response appears to be characterised by three regimes (see Figure~\ref{fig:regions_active}):
(I) at small strains with respect to $\bar{\l}_0$, with $\bar{\l}_0$ such that $\tau^{0,\fibre}_{xx}(\bar{\l}_0)=0$, the stress grows nonlinearly with a transition between two linear regimes, similar to that observed on the passive tissue;
(II) at intermediate strains, the stress response is characterised by a plateau, with a constant value of the measured stress; 
(III) at larger strains, within the experimental window, the stress decreases linearly with the stretch ratio.

For our model to capture this behaviour, we prescribe a dependence of the fitting stress--strain function on the elastic strain $\l-\bar{\l}$ of the form~\eqref{eq:tau_xx_fit_tensile_1}-\eqref{eq:tau_xx_fit_tensile_2}, already assumed for the passive response, and we postulate a piecewise linear dependence of the fibre elastically-relaxed strain $\bar{\l}$ on the measured stretch $\l$. Specifically, on the one hand, we define the following fitting stress--strain function for the active component:
\begin{subequations}\label{eq:tau_xx_fit_active}
\beq{eq:tau_xx_fit_active_1}
  \tau_{\fit}^{\fibre}(\l)=a k^\fibre(\l-\bar{\l})(\l-\bar\l)
\eeq
with
\beq{eq:tau_xx_fit_active_2}
  k^\fibre(\l-\bar{\l})=
  m^\fibre_0+\frac{r^\fibre}{s^\fibre}\frac{m^\fibre_t-m^\fibre_0}{\l-\bar{\l}}\log\left(\frac{e^{s^\fibre}+e^{\tfrac{s^\fibre}{r^\fibre}(\l-\bar{\l})}}{e^{s^\fibre}+1}\right),
\eeq
\end{subequations}
where $m^\fibre_0,m^\fibre_t,r^\fibre,s^\fibre$ are real parameters. 
In particular, $\tau_{\fit}^{\fibre}(\l)\sim m^\fibre_t(\l-\bar{\l})$ for sufficiently large $\l-\bar{\l}$.
On the other hand, for each of the three strain regions, $\text{I}=[\bar{\l}_0,\bar{\l}_1]$, $\text{II}=[\bar{\l}_1,\bar{\l}_2]$, and $\text{III}=[\bar{\l}_2,\bar{\l}_3]$, identified by the fitting parameters $\bar{\l}_j$, we postulate
\beq{eq:active_lambda_plastic}
  \bar\l=h_j\l+q_j \quad \text{for }\l\in[\bar{\l}_{j-1},\bar{\l}_j],
\eeq
for $j=1,2,3$, with $h_1=0$ and $h_2=1$.
The constants $q_j$ are determined in terms of the fitting parameters, $\bar{\l}_0$, $\bar{\l}_1$, $\bar{\l}_2$ and $h_3$, by imposing $\tau_{\fit}^{\fibre}(\bar{\l}_0)=0$ in I and the continuity of $\tau_{\fit}^{\fibre}(\l)$ at $\bar{\l}_1$ and $\bar{\l}_2$.

The constitutive dependence~\eqref{eq:active_lambda_plastic} of $\bar{\l}$ on $\l$ is a phenomenological relation, calibrated to reproduce the three regimes of the active response shown in Figure~\ref{fig:regions_active}. As such, it is not derived from micro-mechanical arguments.
It may nevertheless be formally viewed as reminiscent of a quasi-static plastic flow: within each linear branch, the incremental change in the relaxed fibre stretch is proportional to that of the measured stretch $\delta\bar\l=h_j\,\delta\l$ for $\l\in[\bar{\l}_{j-1},\bar{\l}_j]$. The analogy, however, does not imply $\delta\bar\l$ is an evolving internal state variable in the sense of plasticity, which would require history-dependence and irreversibility. We rather interpret the presence of these three regimes as follows. There exists a minimum relaxed length for the fibres under activation that cannot be further reduced, and this is reached in region I. When the strain is above a certain level, the strength of activation is only able to counteract a maximum value of applied stress, giving rise to the plateau in region II. Eventually, when the overall strain is sufficiently large, as in region III, the activation capabilities are hindered and we observe a decreasing level of active stress.

Ultimately, the active part of the model is fully characterised by eight material parameters, to be determined through the fitting function~\eqref{eq:tau_xx_fit_active} with the experimental data set under condition~\eqref{eq:active_lambda_plastic}: $\bar{\l}_0$, $\bar{\l}_1$, $\bar{\l}_2$, $m^\fibre_0$, $m^\fibre_t$, $r^\fibre$, $s^\fibre$, $h_3$. Table~\ref{tab:par_active} collects the values of such parameters and the corresponding uncertainties obtained through the fitting procedure on the reference experimental data by~\cite{hawkins1994comprehensive}.

\begin{table}[htb]
\caption{Values of the model parameters entering the fitting functions~\eqref{eq:tau_xx_fit_active} (left) and ~\eqref{eq:tau_xx_fit_tensile} (right) for the experimental data set of~\cite{hawkins1994comprehensive}.}
\centering
\begin{tabular}[t]{c c c} 
 \toprule
 Parameter & Value & Standard error\\ [0.5ex] 
 \midrule
 $\bar{\l}_0$ & -0.39 & 0.07\\
 $\bar{\l}_1$ & -0.07 & 0.01\\
 $\bar{\l}_2$ & 0.27 & 0.02\\
 $m^\fibre_0$ [kPa] & 46 & 7\\
 $m^\fibre_t$ [kPa] & 230 & 6\\
 $r^\fibre$ & 0.07 & 0.01\\
 $s^\fibre$ & 7 & 4\\
 $h_3$ [kPa] & -227 & 9\\[0.5ex]
 \bottomrule
\end{tabular}\hspace{4mm}
\begin{tabular}[t]{c c c} 
 \toprule
 Parameter & Value & Standard error\\ [0.5ex] 
 \midrule
 $m^\ecm_0$ [kPa] & 10 & 1\\
 $m^\ecm_t$ [kPa] & $1.2\cdot10^3$ & $1\cdot10^2$\\
 $r^\ecm_0$ & 0.305 & 0.001\\
 $s^\ecm_{01}$ & 18 & 1\\[0.5ex]
 \bottomrule
\end{tabular}
\label{tab:par_active}
\end{table}

The functional dependence of the material function $c_{\fibre,1}$ on the invariant $\l-\bar{\l}$ is then obtained setting $\eta(\l)\tau_{xx}^{0,\fibre}(\l)=\tau_{\fit}^{\fibre}(\l)$. This gives $\hat{c}_{\fibre,1}(\l-\bar{\l})=a k^\fibre(\l-\bar{\l})$, or, equivalently,
\beq{eq:material_functions_c_active}
  \hat{c}_{\fibre,1}(\l_{\fibre,1})=a k^\fibre\Bigl(\sqrt{\tfrac23}\l_{\fibre,1}\Bigr).
\eeq
Ultimately, from~\eqref{eq:stress_incompressible_transverse},~\eqref{eq:hyp_7}, and~\eqref{eq:material_functions_c_active}, we arrive at the following constitutive law for the fibre component:
\beq{eq:model_active}
    \hat{\*\sigma}_\fibre(\*\l_\fibre)=
    \hat{c}_{\fibre,1}(\hat{\rho}_{\fibre,1}(\*\l_\fibre))\!\mx{\l_{\fibre,1}\\0\\0\\0\\0}.
\eeq

\begin{figure}[htb]
    \centering
    \includegraphics[width=0.9\textwidth]{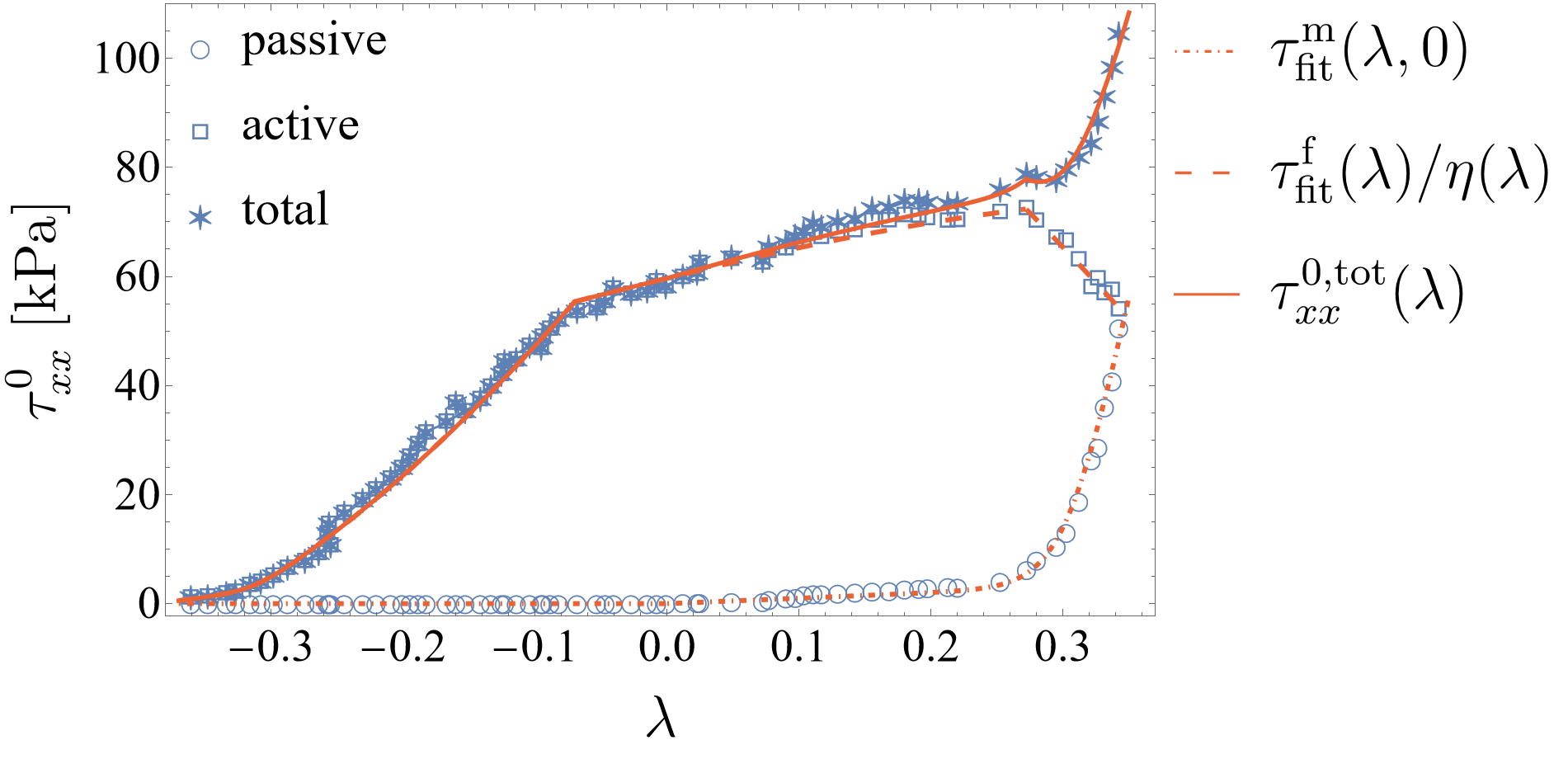}
    \caption{Fits of the experimental stress--strain data sets (uniaxial along-fibre tensile tests for passive and tetanised muscle tissue) of~\cite{hawkins1994comprehensive} with the proposed model~\eqref{eq:model_final}. The fitting function for the total stress, in the presence of activation, (filled line) is obtained by superimposing the passive (matrix) response (dotted line) to the active (fibre) response (dashed line).}
    \label{fig:fit_active}
\end{figure}

The total stress field is finally obtained by adding the fibre and matrix fields. Hence, the constitutive law of the model that we propose is
\beq{eq:model_final}
    \hat{\*\sigma}_\tot(\*\l_\ecm,\*\l_\fibre)=
    \!\mx{
    \hat{c}_{\ecm,1}(\hat{\rho}_\ecm(\*\l_\ecm))\l_{\ecm,1}
    +\hat{c}_{\fibre,1}(\hat{\rho}_{\fibre,1}(\*\l_\fibre))\l_{\fibre,1}
    \\
    \hat{c}_{\ecm,2}(\hat{\rho}_\ecm(\*\l_\ecm))\l_{\ecm,2}
    \\
    \hat{c}_{\ecm,3}(\hat{\rho}_\ecm(\*\l_\ecm))\l_{\ecm,3}
    \\
    \hat{c}_{\ecm,3}(\hat{\rho}_\ecm(\*\l_\ecm))\l_{\ecm,4}
    \\
    \hat{c}_{\ecm,2}(\hat{\rho}_\ecm(\*\l_\ecm))\l_{\ecm,5}
    }.
\eeq
Figure~\ref{fig:fit_active} shows the fitting of the experimental data sets of the passive and active measured stresses as functions of the along-fibre stretch $\l$ with the prescribed functions $\tau_{xx}^{0,\ecm}$ and $\tau_{xx}^{0,\fibre}$, respectively, with material parameters given in Table~\ref{tab:par_active}. The measured total stress is well interpolated by the function $\tau_{xx}^{0,\tot}(\l)$ resulting from our model.

\section{Conclusions}
\label{sec:conclusions}
In this work, we proposed a constitutive model for the passive and active quasi-static responses of skeletal muscle tissue in the setting of nonlinear Cauchy elasticity, following the framework recently developed by \citet{galasso2025adapted}. 
The present framework departs from classical hyperelastic approaches by prescribing the constitutive relation directly, without requiring it to derive from an underlying strain energy. One key advantage is that the functional form of each stress component can be identified independently and directly from the experimental stress--strain data, rather than being constrained by the structure of a pre-assigned energy function.
Relying on the stress--strain measurements of \cite{takaza2013anisotropic} and \cite{hawkins1994comprehensive}, we view the muscle as a composition of two transversely isotropic and incompressible materials, the mechanical contributions of which depend on activation. The latter is introduced as a binary control parameter, which affects their elastically-relaxed configurations. We assumed an additive decomposition of the total stress as a function of both logarithmic elastic strains into a \emph{matrix} component and a \emph{fibre} component. The matrix component is responsible for the passive behaviour and its response to strain is independent of the activation state, while---at nonzero activation parameter---the fibre component accounts for an active stress, which depends on a non-trivial elastically-relaxed state. 

The constitutive prescription for the passive (matrix) response is characterised in our model by three material functions, representing the deviatoric along-fibre stress, the cross-fibre normal and shear stresses, and the along-fibre shear stress components, respectively. We postulated that each is a nonlinear function of the invariant given by the norm of the (logarithmic) elastic strain. 
Additional contributions to the stress components, allowed by the general framework that we employ, were set to zero in an effort to keep the model as simple as possible. As a matter of fact, we find that the components that we consider are sufficient to describe the material behaviour. The model parameters are determined, for the passive behaviour, on the basis of a subset of the available data, while the remaining data are used to validate the model, with a very satisfactory result. 
As noted above, prescribing each stress component independently allowed us to capture features of the passive response that a single-potential formulation would not easily accommodate. In particular, we interpreted the along-fibre stiffening trend as reflecting the progressive recruitment of the collagen network at increasing strain, with an increasing portion of the network engaging in the elastic response as the tissue is stretched in the fibre direction, while the extracellular matrix contributes immediately to the response when stretched orthogonally to the fibres. This is consistent with the observed independence of tissue stiffness from fibre orientation at large strains, indicating that the extracellular matrix becomes the dominant contributor in that regime. Capturing this orientation- and strain-dependent transition directly from an invariant-based energy function would require additional structural assumptions in a hyperelastic formulation. As a further practical advantage, the resulting material parameters were, in most cases, identified through linear regression, rather than the nonlinear optimisation typically required for strain-energy-based models.

In the presence of activation, we adopted an active strain approach for the active (fibre) stress component. We assumed a response that depends on the elastic component of the strain, coupled with a variable elastically-relaxed state. We assumed that the fibre component contracts in the longitudinal direction when the activation control parameter is switched on and exerts a deviatoric normal stress in the same direction.
The fitting of the material parameters relative to the active component of the stress greatly benefits from the identification of three clearly distinct regimes, from which we can highlight the piecewise-linear dependence of the elastically-relaxed state of the fibre on the deformation of the matrix, related to a saturation of the fibre component capability of sustaining traction.

Overall, we obtained a quite complete model for the mechanics of skeletal muscles subject to homogeneous (uniaxial) quasi-static deformations.
We plan to use this model for a computational investigation of the mechanics of three-dimensional muscles with more realistic geometries, thereby analysing the performance of the model in the presence of non-homogeneous deformations. Within such a study, it will be important to consider the presence of a dynamical activation process. This will require adding an evolution equation or a controlled time-dependence of the activation parameter and exploring its effect on the evolution of the elastically-relaxed state of the fibre component. More generally, when extending the framework to dynamic loading and activation, rate- and history-dependent effects could also be incorporated. On the one hand, it would be essential to include dissipative viscosity in the matrix component, consistent with the well-documented viscoelastic behaviour of collagenous tissue. On the other hand, for the fibre component, one might expect viscous dissipation to be negligible with respect to the activation process, which occurs on a shorter time scale; even so, viscoelasticity might still be important to account for history-dependent phenomena such as fatigue.
The present framework could easily accommodate these effects: the components of the stress tensor in the adapted basis remain unchanged, and thus rate-dependence would enter through the dependence of the same material functions on a larger set of variables. Consistent with the overall approach adopted in this work, the specific form of these contributions should be identified directly from dynamic experimental data, rather than prescribed a priori.

\section*{CRediT author statement}

\textbf{Sara Galasso:} Conceptualization, Methodology, Formal analysis, Data Curation, Writing - Original Draft, Writing - Review \& Editing, Visualization.

\textbf{Giulio G.\ Giusteri:}
Conceptualization, Writing - Review \& Editing, Supervision, Funding acquisition.

\section*{Declaration of competing interests}
The Authors declare no competing interests.

\section*{Acknowledgements}
Project funded by the European Union–Next Generation EU under the National Recovery and Resilience Plan (NRRP),
Mission 4 Component 2 Investment 1.1–Call PRIN 2022 of Italian Ministry of University and Research Project no. 202249PF73
“Mathematical models for viscoelastic biological matter.”

\bibliographystyle{elsarticle-harv}
\bibliography{biblio.bib}

\end{document}